\documentclass[fleqn,usenatbib]{mnras}

\usepackage[T1]{fontenc}
\usepackage{ae,aecompl}
\usepackage[utf8]{inputenc}
\usepackage{float}
\usepackage{amsmath,amssymb}
\usepackage{mathrsfs}
\usepackage{theorem}
\usepackage{graphicx}
\usepackage{color}
\usepackage{wasysym}
\usepackage{hyperref} 
\usepackage{verbatim}
\usepackage{tabularx}

\definecolor{blue}{rgb}{0,0,1}
\definecolor{green}{rgb}{0,0.65,0.5}
\definecolor{verde}{rgb}{0.,.5,0.5}
\definecolor{marron}{rgb}{0.7,0.2,0.1}
\definecolor{red}{rgb}{1,0,0}
\definecolor{vio}{rgb}{1,0,1}
\definecolor{ama}{rgb}{1,1,0}

\title[Synthetic gravitational lens image of...]
{\bf Synthetic gravitational lens image of the Sagittarius A${}^*$ black hole with a thin disk model }

\author[E.F.Boero and O.M.Moreschi]{
Ezequiel F. Boero,$^{1}$\href{https://orcid.org/0000-0001-5877-3565}{\includegraphics[scale=0.4]{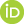}}\thanks{E-mail: ezequiel.boero@unc.edu.ar} 
and
Osvaldo M. Moreschi\href{https://orcid.org/0000-0001-9753-3820}{\includegraphics[scale=0.4]{ORCIDID_icon24x24.png}},$^{2,3}$\thanks{E-mail: o.moreschi@unc.edu.ar}
	\\
	{\rm \small $^{1}$ Instituto de Astronom\'\i{}a Te\'{o}rica y Experimental (IATE), CONICET,} \\
	{\rm \small Observatorio Astron\'{o}mico de Córdoba,}\\
	{\rm \small Laprida 854, (X5000BGR) C\'{o}rdoba, Argentina.}\\
	{\rm \small $^{2}$Facultad de Matem\'{a}tica, Astronom\'\i{}a, F\'\i{}sica y Computaci\'{o}n (FaMAF),}\\
	{\rm \small Universidad Nacional de C\'{o}rdoba,} \\ 
	{\rm \small $^{3}$Instituto de F\'\i{}sica Enrique Gaviola, IFEG, CONICET, } \\
	{\rm \small Ciudad Universitaria, (5000) C\'{o}rdoba, Argentina.} 
}

\date{Accepted XXX. Received YYY; in original form ZZZ}

\begin{document}
	\label{firstpage}
	\pagerange{\pageref{firstpage}--\pageref{lastpage}}
	\maketitle
	
\begin{abstract}
The images of Sagittarius A${}^*$ published by the Event Horizon Telescope (ETH) Collaboration in 2022
present features that were associated with an emission ring consistent with what is expected from 
an accretion disc surrounding the supermassive black hole at the center of our Galaxy. 
Here, we generate images of Sgr~A${}^*$ across different
configurations of a simple accretion disc model 
that became successful, in our previous work,
in reproducing the main features 
observed in M87*. 
Their best image, here reproduced in Fig. \ref{fig:eht-image},
suggests a geometric configuration of an inclined disk
with three bright regions; which we have considered as our
first configuration.
Since we were not convinced with the results of this first
configuration, we also
explore in detail the case of nearly edge-on 
orientations which are a priori
the expected geometry for a relaxed disc,
as seen from the plane of the galaxy. 
We have produced simulated images using an efficient 
ray tracing and geodesic deviation methodology that allows to
account for deformation, relativistic and magnification
effects.
We compare our synthetic images with the EHT
images reconstructed with data 
from April 6 and 7 of 2017.
We found that, although the EHT Collaboration seems to
discard the image from April 6, our best suggested image
resembles the output from the {\sc Themis} pipeline
for April 6; which for us gives support for
the edge-on configuration.

\end{abstract}


\section{Introduction}\label{sec:Introduction}

\begin{figure}
\centering
\includegraphics[clip,width=0.4\textwidth]{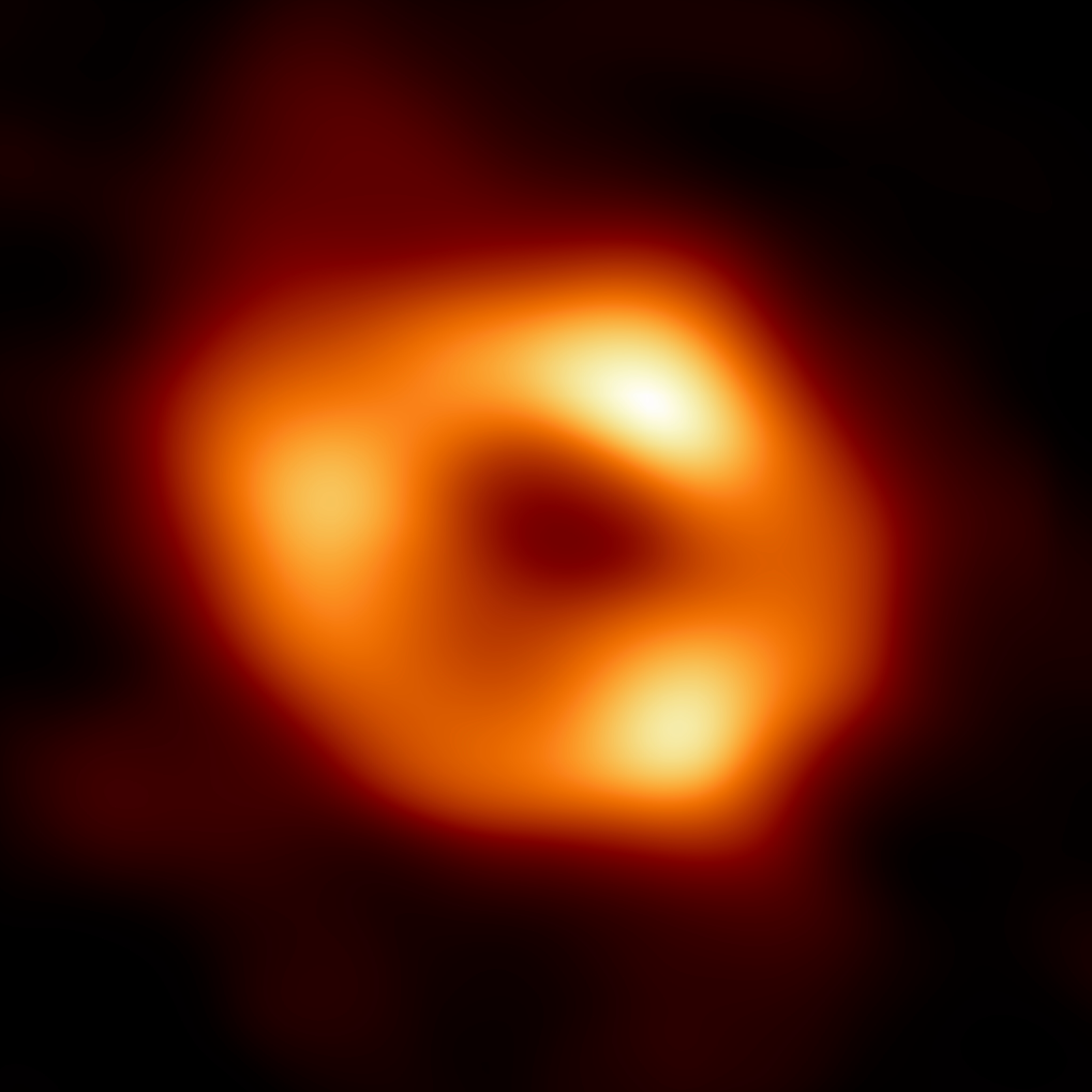}
\caption{Image from the EHT 
Collaboration paper III \protect\cite{EventHorizonTelescope:2022wok} 
of Sgr~A${}^*$;
where they show an average of 
reconstructed images for April 7,
in their Figure 13.
}
\label{fig:eht-image}
\end{figure}

The immediate surroundings of supermassive black holes (SMBH) in scales of a few horizon radius 
were resolved for the first time in recent years \citep{Akiyama:2019cqa, EventHorizonTelescope:2022xnr}.
The observed systems were M87* located at the center of the nearby elliptical galaxy M87\citep{Akiyama:2019bqs} 
and Sgr~A${}^*$
\citep{EventHorizonTelescope:2022xnr, EventHorizonTelescope:2022vjs, EventHorizonTelescope:2022wok, EventHorizonTelescope:2022exc, EventHorizonTelescope:2022urf, EventHorizonTelescope:2022xqj,
EventHorizonTelescope:2022ago,
EventHorizonTelescope:2022okn,
EventHorizonTelescope:2022ppi} 
the SMBH hosted in our Galaxy.
Their very compact nature together with the properties of emission from the innermost parts required
the coordination of the array of antennas comprised in the Event Horizon Telescope (EHT) in order to
achieve the necessary precision to image these objects. 
The EHT is an array of Very Long Baseline Interferometry (VLBI)
with the size of the order of the Earth diameter that at
millimetre and shorter wavelengths is able to detect
angular scales of a few $\mu$arcsec.
It observes in the millimetre radio wave band ($\lambda \sim 1.3$mm),
where the emission coming from the accretion disc near the event horizon 
becomes transparent to the material in its neighbourhood.
The images obtained in this manner suffer the
restrictions of the VLBI method,
due to the intrinsic limitations of the observational setting, which involves only a limited number of points in Fourier space.
As a consequence the process which allows to reconstruct an image of the system 
becomes rather sophisticated and complex since it 
necessarily involves the use of model assumptions. 
In particular, several methods \citep{EventHorizonTelescope:2022wok, Akiyama:2019bqs} are used to infer images from the measured visibilities in radio which present some expected differences among them.
Then, the crucial question is to what extent intrinsic limitations in the imaging process
allow to extract a clear interpretation of the diverse features of the images.
The answer is not completely clear even though the EHT Collaboration has presented evidence supporting some common features sheared in numerous image results.
Indeed, most of the reconstructed images obtained with different methods often reveal a prominent ring-like structure, which would correspond to the emission from hot gas around the event horizon of the black holes as expected on physical grounds.
Besides, comparison with a template bank of GRMHD numerical simulations indicates that the geometry of the central object is compatible with a Kerr black hole(BH) \citep{EventHorizonTelescope:2022xqj}.
Nevertheless, the degree of degeneracy in the models that one can fit to the images
makes difficult to give a definite answer.
In particular, concerning to the representative image released by the EHT Collaboration reproduced in Fig. \ref{fig:eht-image}, one notices the presence of three bright spots
inside a little distorted disk of emission.
The presence of these three bright spots is not a feature present in every reconstruction,
with different pipelines, but instead it appears with different degree of occurrence 
as mentioned in \cite{EventHorizonTelescope:2022wok}.
This seems to be particularly complicated
since Sgr~A* shows time variability \citep{EventHorizonTelescope:2022xnr} on the order
of minutes \citep{Johannsen:2015hib} .

We have studied in the past the case of M87* \cite{Boero:2021} assuming a simple scenario for the 
emission in terms a disc region having two temperatures.
Such kind of models assume a collisionless plasma
where the electron distribution function is approximated by a thermal distribution (see for example \cite{Leung2011ApJ, Bandyopadhyay:2021jgh, EventHorizonTelescope:2022urf}).
These seem to be reasonable assumptions in order to account some observational aspects of the emission spectra at radio and X-ray wavelengths \citep{Yuan:2001dm, Yuan:2003dc}. 
Using a new approach that combines the 
integration of geodesics and geodesic deviation equation in a very efficient way \cite{Boero:2019zkq} we obtained images with
this prescription.
As a result we found that our images could fit very well the reported final images of the 
EHT Collaboration.

In this article we focus on the most recent EHT 
publications on Sgr~A${}^*$\cite{EventHorizonTelescope:2022xnr, EventHorizonTelescope:2022vjs, EventHorizonTelescope:2022wok, EventHorizonTelescope:2022exc, EventHorizonTelescope:2022urf, EventHorizonTelescope:2022xqj};
and
we carry a similar analysis to our previous work,
motivated by the good results we had obtained;
and noting that both SMBH have almost the same
angular size.

For the case of M87* and Sgr~A${}^*$ the discs is expected to present properties consistent with emission 
coming from an optically thin disc where very hot flows are accreted mainly through advection processes
\citep{Shapiro:1976fr, Ichimaru:1977uf, Yuan:2014gma}.
The material in the vicinity of the BH is an ionized plasma with electrons and ions having
different temperatures.

The EHT Collaboration has provided a physical scenario for the geometry and the accretion flow orbiting 
Sgr A${}^*$ based on comparison of the data with several synthetic images from diverse models. 
From their analysis they claim that the orbiting plasma is most probably described by magnetically arrested 
disc (MAD) which tend to be strongly magnetized.

As in their previous work on M87, the EHT team
emphasizes that the images could be mainly
understood in terms of a ring structure;
however we have also shown in our article
on the same system, that the images might
as well be understood in terms of the
natural structure of a disk model.
What is somehow striking in the described
construction process used by EHT Collaboration
in the 
Sgr~A${}^*$\citep{EventHorizonTelescope:2022wok}
case is the fact, see for example
their figures 7, 11 and 12,
that the models involving rings,
crescents, disks
and other 
general relativistic 
magnetohydrodynamic (GRMHD)
calculations,
are almost face-on configurations.
Since as observers we are situated more or less
in the equatorial plane of the galaxy,
it would be natural to assume first
a ring or disk structure that it would
be close to its plane of symmetry, 
that is, one would expect to have
an edge-on view.
We tackle this issue in this article.

One of the main characteristics of the observation of Sgr~A${}^*$
is its time variability. In turn, this complicates considerably
the reconstruction of an image from the observed data.
For this reason as mentioned in \citep{EventHorizonTelescope:2022wok, EventHorizonTelescope:2022exc} the EHT team
must rely on different kind of models.
In fact, their methodology for the construction of images is to use
a variety of models for the matter distribution
around the supermassive black hole.

Due to the unexpected geometry suggested by the
EHT image of reference \citep{EventHorizonTelescope:2022wok}, 
here shown in Fig. \ref{fig:eht-image},
we study first a thin accretion disk whose plane
is inclined with respect of the plane of the galaxy,
as suggested by the EHT image.
We have considered a big range for the projected angular 
momentum of the black hole with the line of sight,
and also a range of values for the possible total angular
momentum of the black hole.
Since we were not convinced with the results of this study,
that we present below, we have also considered the
more natural assumption of an accretion disk and
black hole angular momentum that have small angles
with respect to the plane of the  galaxy and
angular momentum of the galaxy respectively.
In our opinion, this last model gives better results
than the previous one.

Whenever possible, for tensors and vectors objects,
we will employ standard abstract index notation 
with Latin letters $a,b,c,...$ for tensor fields. 
Our choice of signature for the spacetime metric $g_{ab}$ is $(+,-,-,-)$.

The organization of the article is as follows.
In the next second section we review the basic equations
that we use in our construction. Since they were discussed
at length in our previous article \citep{Boero:2021}
we present here a summary of the basic general dynamical
equations for the deviation vector and the optical scalars
equations.
In section \ref{sec:Null+geod+Kerr}
we describe the geometry, the evolution equations,
the conserved quantities, the initial conditions
and details of the matter model.
Our image constructions are presented in section
\ref{sec:Results}.
We take the opportunity in this case to also present
the calculation of the expected silhouette of the
black hole for different configurations in
section \ref{sec:silueta}.
We reserve the final section \ref{sec:Final+commnets}
for some recapitulation and comments;
where we argue that our resulting image is
very closely related to the April 6 image
coming from the their {\sc Themis} pipeline.

\section{Exact gravitational lens optical scalars}\label{sec:Exact+grav+lens+opt+scalars}

\subsection{The basic equations}

Although we have discussed the notion of `exact gravitational lens optical scalars'
in the past\citep{Boero:2021}, we here introduce an alternative for the initial conditions,
improve the notation
and correct a typo of the last version.

The general setting is to consider the null geodesics in the past null cone of an
observer moving with 4-velocity  $v^a$. The tangent vectors to the null geodesics
are denoted by $\ell^a$, which are past directed; so that are normalized
by the condition:
\begin{equation}\label{eq:ell+v+minus1}
\ell^a v_a = -1,
\end{equation}
and since they are null geodesics, we also have
\begin{equation}\label{eq:ell+nabla+ell}
\ell^a \nabla_a \ell^b = 0,
\end{equation}
where $\nabla_a$ is the covariant derivative in terms of abstract indices.

In order to calculate the optical scalars we need to introduce the geodesic deviation vector 
$\varsigma^a$,  in terms of a suitable null tetrad $(\ell^a, m^a ,\bar{m}^a, n^a)$ where $m^a$ 
and $\bar{m}^a$ is a pair of complex conjugated vectors and $n^a$ an additional 
real null vector.
We choose the two complex vectors $m^a$ and $\bar{m}^a$ to be parallel 
propagated along the geodesic.
Then, we express the geodesic deviation vector as
\begin{equation}\label{eq:varsigma}
\varsigma^a = \varsigma \bar{m}^a + \bar{\varsigma} m^a + \eta \ell^a.
\end{equation}

Using that $\varsigma^a$ must be Lie transported along the null geodesics, one arrives
at the geodesic deviation equations:
\begin{equation}\label{eq:geod+dev+equation}
\ell^a \nabla_a \left( \ell^b \nabla_b \varsigma^d \right) = 
R_{abc}^{\;\;\;\;\;d}\ell^a \varsigma^b \ell^c;	
\end{equation}
where $R_{abc}^{\;\;\;\;\;d}$ denotes the Riemann tensor.
Then the task is to solve the couple system of differential equations 
\eqref{eq:ell+nabla+ell} and \eqref{eq:geod+dev+equation}.

\subsection{Optical scalars}

Let us define here explicitly the exact gravitational optical scalars.

The situation we are considering is that of an observed deviation $\delta \theta^i$ 
from a reference observed direction in the sky, denoted by $\theta$, corresponding to the direction of the
null geodesic described by $\ell^a$.
In the absence of gravitational effects one would detect the deviation  $\delta \beta^i$
with respect to the reference direction $\beta$.
Then, 
the optical scalars $\kappa$, $\gamma_{1}$, $\gamma_{2}$ and $\omega$ are defined 
through the linear relation between $\delta \theta^i$ and 
$\delta \beta^i$
\begin{equation}\label{eq:linear+relation+angles}
\delta \beta^i = \mathcal{A}^i_{\;\,j} \, \delta \theta^j,
\end{equation}
were the matrix $\mathcal{A}^i_{\;\,j}$ is given by
\begin{equation}\label{eq:matrizA}
\mathcal{A}^i_{\;\,j} = \begin{pmatrix}
1 - \kappa - \gamma_1 & -\gamma_2 - \omega \\
-\gamma_2 + \omega    & 1 - \kappa + \gamma_1
\end{pmatrix}.
\end{equation}

Let us note that the component $\varsigma$ is complex,
and that their real and imaginary parts $(\varsigma_{R},\varsigma_{I})$, 
can be used as components in a real 2-dimensional plane;
orthogonal to $\ell^a$ and therefore they will be related to the
angles $\delta \theta^i$ and $\delta \beta^i$.
We will denote with $\upsilon_{R}$ and $\upsilon_{I}$ the derivatives
of $\varsigma_{R}$ and $\varsigma_{I}$ respectively, along the null geodesic $\ell^a$.
Then the initial conditions for the couple differential equation,
that is at the observer position, are:
\begin{equation}
\delta \theta^i \equiv 
\left. 
\begin{pmatrix}
\upsilon_{R}^{} \\
\upsilon_{I}^{}
\end{pmatrix}
\right|_{\lambda=0}
\equiv 
\left.
\begin{pmatrix}
\ell \left(\varsigma_{R}^{}\right) \\
\ell\left( \varsigma_{I}^{} \right)
\end{pmatrix}
\right|_{\lambda=0}
,
\end{equation}
and
\begin{equation}
\left. \varsigma^i \right|_{\lambda = 0} 
= 
\left.
\begin{pmatrix}
\varsigma_{R}^{} \\
\varsigma_{I}^{}
\end{pmatrix}
\right|_{\lambda = 0}
= 
\begin{pmatrix}
0 \\
0
\end{pmatrix}
;
\end{equation}
where we are using $\lambda$ to denote the affine parameter along the null geodesic.

At the target point the affine parameter has the value $\lambda_s$
and then by definition one has
\begin{equation}
\delta \beta^i \equiv \frac{1}{\lambda_s} 
\begin{pmatrix}
\varsigma_{R}^{} \\
\varsigma_{I}^{}
\end{pmatrix}
\bigg|_{\lambda = \lambda_s}
.
\end{equation}
Let us note that $\varsigma$ has units of length.
The unit of length might be relatively big or small with
respect to the size of the system.
Then, equations \eqref{eq:linear+relation+angles} and \eqref{eq:matrizA} explicitly become:
\begin{align}
\frac{\varsigma_{R}^{}}{\lambda_s}	 =& \big( 1 - \kappa - \gamma_1 \big) \upsilon_{R}^{}   - 
\big(\gamma_2 + \omega \big) \upsilon_{I}^{} , \\
\frac{\varsigma_{I}^{}}{\lambda_s}	 =& - \big(\gamma_2 - \omega \big) \upsilon_{R}^{}  
+ \big(1 - \kappa + \gamma_1 \big) \upsilon_{I}^{} . 
\end{align}
In relation to the integration process of the geodesic deviation equation,
it is probably worthwhile to mention that while on the right hand side
of the above equations, the $\upsilon_{R}$ and $\upsilon_{I}$ are evaluated
at the observer position, on the left hand side the $\varsigma_{R}$ and
$\varsigma_{I}$ are evaluated at the source position.
Since it is a 2-dimensional system we need to consider two independent
initial conditions.
Due to the fact that we are in the galaxy and that a possible disk around
Sgr~A${}^*$ would probably sustain a very small angle,
we have considered in the numerical calculation a couple of initial
conditions; just to see if the computation was sensible to them. 

The first standard choice for initial conditions is to take
\begin{align}
\left. 
\begin{pmatrix}
\varsigma_{R_1}^{} \\
\varsigma_{I_1}^{}
\end{pmatrix}
\right|_{\lambda=0}
=&
\begin{pmatrix}
0 \\
0
\end{pmatrix}
\label{eq:IC_firstchoie_devgeod_0}
,
\\
\left. 
\begin{pmatrix}
\upsilon_{R_1}^{} \\
\upsilon_{I_1}^{}
\end{pmatrix}
\right|_{\lambda=0}
=&
\frac{1 [\mathsf{L}]}{\lambda_s}
\begin{pmatrix}
1 \\
0
\end{pmatrix}
\label{eq:IC_firstchoie_devgeod_1}
,
\end{align}
and
\begin{align}
\left. 
\begin{pmatrix}
\varsigma_{R_2}^{} \\
\varsigma_{I_2}^{}
\end{pmatrix}
\right|_{\lambda=0}
=&
\begin{pmatrix}
0 \\
0
\end{pmatrix}
\label{eq:IC_firstchoie_devgeod_2}
,
\\
\left. 
\begin{pmatrix}
\upsilon_{R_2}^{} \\
\upsilon_{I_2}^{}
\end{pmatrix}
\right|_{\lambda=0}
=&
\frac{1 [\mathsf{L}]}{\lambda_s}
\begin{pmatrix}
0 \\
1
\end{pmatrix}
\label{eq:IC_firstchoie_devgeod_3}
,
\end{align}
where $\mathsf{L}$ is the unit of length 
(which it could be taken as $\lambda_s$)
and
which yields the final linear system of equations for the quantities
$\kappa$, $\gamma_{1}$, $\gamma_{2}$ and $\omega$:
\begin{align}
\frac{\varsigma_{R1}^{}}{[\mathsf{L}]}
=& 
\big( 1 - \kappa - \gamma_1 \big)  , \\
\frac{\varsigma_{I1}^{}}{[\mathsf{L}]}
=& - \big(\gamma_2 - \omega \big) , \\
\frac{\varsigma_{R2}^{}}{[\mathsf{L}]}
=& -\big(\gamma_2 + \omega \big) , \\
\frac{\varsigma_{I2}^{}}{[\mathsf{L}]}
=& \big( 1 - \kappa + \gamma_1 \big) ,
\end{align}
or equivalently 
\begin{align}
\kappa =& 1 - \frac{\varsigma_{R1}^{} + \varsigma_{I2}^{}}{2 [\mathsf{L}]} , \\
\gamma_1 =& \frac{\varsigma_{I2}^{} - \varsigma_{R1}^{}}{2 [\mathsf{L}]}, \\
\gamma_2 =& - \frac{\varsigma_{I1}^{} + \varsigma_{R2}^{}}{2 [\mathsf{L}]}, \\
\omega =& \frac{\varsigma_{I1}^{} - \varsigma_{R2}^{}}{2 [\mathsf{L}]}.
\end{align}

It is worthwhile to mention that in a numerical work the choice of the unit
of length $[\mathsf{L}]$ is an important detail to have in mind, since;
although the geodesic deviation equation is linear, and therefore we would
be free to use any magnitudes, from the numerical point of view,
one does not want to have very different orders of magnitude for the couple
system of ordinary differential equations. This is precisely
adjusted with the choice of unit of length.
It is for this reason that we have chosen now to introduce explicitly
the choice of unit of length in the notation.

The second set of initial conditions we have considered is just a 45º rotation
of the first one; namely, we take at the observer position:
\begin{align}
\left. 
\begin{pmatrix}
\varsigma_{R_1}^{} \\
\varsigma_{I_1}^{}
\end{pmatrix}
\right|_{\lambda=0}
=&
\begin{pmatrix}
0 \\
0
\end{pmatrix} 
\label{eq:IC_secondchoie_devgeod_0}
,
\\
\left. 
\begin{pmatrix}
\upsilon_{R_1}^{} \\
\upsilon_{I_1}^{}
\end{pmatrix}
\right|_{\lambda=0}
=&
\frac{1 [\mathsf{L}]}{\lambda_s}
\begin{pmatrix}
\frac{1}{\sqrt{2}} \\
\frac{1}{\sqrt{2}}
\end{pmatrix}
\label{eq:IC_secondchoie_devgeod_1}
,
\end{align}
and
\begin{align}
\left. 
\begin{pmatrix}
\varsigma_{R_2}^{} \\
\varsigma_{I_2}^{}
\end{pmatrix}
\right|_{\lambda=0}
=&
\begin{pmatrix}
0 \\
0
\end{pmatrix}
\label{eq:IC_secondchoie_devgeod_2}
,
\\
\left. 
\begin{pmatrix}
\upsilon_{R_2}^{} \\
\upsilon_{I_2}^{}
\end{pmatrix}
\right|_{\lambda=0}
=&
\frac{1 [\mathsf{L}]}{\lambda_s}
\begin{pmatrix}
-\frac{1}{\sqrt{2}} \\
\frac{1}{\sqrt{2}}
\end{pmatrix}
\label{eq:IC_secondchoie_devgeod_3}
.
\end{align}
These equations yield the final linear system of equations for the quantities
$\kappa$, $\gamma_{1}$, $\gamma_{2}$ and $\omega$:
\begin{align}
\frac{\varsigma_{R1}^{}}{[\mathsf{L}]}	 =& \big( 1 - \kappa - \gamma_1 \big) \frac{1}{\sqrt{2}}   
- \big(\gamma_2 + \omega \big) \frac{1}{\sqrt{2}} , \\
\frac{\varsigma_{I1}^{}}{[\mathsf{L}]}	 =& 
- \big(\gamma_2 - \omega \big) \frac{1}{\sqrt{2}}  
+ \big(1 - \kappa + \gamma_1 \big) \frac{1}{\sqrt{2}} . 
\end{align}
\begin{align}
\frac{\varsigma_{R2}^{}}{[\mathsf{L}]}	 =& 
-\big( 1 - \kappa - \gamma_1 \big) \frac{1}{\sqrt{2}}   
- \big(\gamma_2 + \omega \big) \frac{1}{\sqrt{2}} , \\
 \frac{\varsigma_{I2}^{}}{[\mathsf{L}]}	 =& 
  \big(\gamma_2 - \omega \big) \frac{1}{\sqrt{2}}  
+ \big(1 - \kappa + \gamma_1 \big) \frac{1}{\sqrt{2}} 
,
\end{align}
or equivalently 
\begin{align}
\kappa + \omega =& 1 - \frac{\varsigma_{R1}^{} + \varsigma_{I2}^{}}{\sqrt{2} [\mathsf{L}]} , \\
\gamma_1 + \gamma_2  =& \frac{\varsigma_{I2}^{} - \varsigma_{R1}^{}}{\sqrt{2} [\mathsf{L}]}, \\
- \gamma_1 + \gamma_2 =& - \frac{\varsigma_{I1}^{} + \varsigma_{R2}^{}}{\sqrt{2} [\mathsf{L}]}, \\
- \kappa + \omega =& -1 + \frac{\varsigma_{I1}^{} - \varsigma_{R2}^{}}{\sqrt{2} [\mathsf{L}]}
;
\end{align}
so that
\begin{align}
\kappa =& 1 +\frac{1}{2}
\big(
 - \frac{\varsigma_{R1}^{} + \varsigma_{I2}^{}}{\sqrt{2} [\mathsf{L}]}
 -\frac{\varsigma_{I1}^{} - \varsigma_{R2}^{}}{\sqrt{2} [\mathsf{L}]}
 \big) , \\
\gamma_1  =& 
\frac{1}{2}
\big(
\frac{\varsigma_{I2}^{} - \varsigma_{R1}^{}}{\sqrt{2} [\mathsf{L}]}
+ \frac{\varsigma_{I1}^{} + \varsigma_{R2}^{}}{\sqrt{2} [\mathsf{L}]}
 \big), \\
 \gamma_2 =& 
\frac{1}{2}
\big(
\frac{\varsigma_{I2}^{} - \varsigma_{R1}^{}}{\sqrt{2} [\mathsf{L}]}
- \frac{\varsigma_{I1}^{} + \varsigma_{R2}^{}}{\sqrt{2} [\mathsf{L}]}
 \big), \\
 \omega =& 
 \frac{1}{2}
 \big(
 \frac{\varsigma_{I1}^{} - \varsigma_{R2}^{}}{\sqrt{2} [\mathsf{L}]}
- \frac{\varsigma_{R1}^{} + \varsigma_{I2}^{}}{\sqrt{2} [\mathsf{L}]}
\big).
\end{align}

It important to note that the calculation of the 
magnification factor $\mu$, in terms of the final values of the deviation vector,
is independent of the initial angle choice and is given by:
\begin{equation}\label{eq:magnification}
\mu = \frac{1}{\left(1 - \kappa \right)^2 - \left(\gamma_1^2 + \gamma_2^2 \right) + \omega^2}
=
\frac{[\mathsf{L}]}{ \varsigma_{I2}^{} \varsigma_{R1}^{} - \varsigma_{I1}^{} \varsigma_{R2}^{}}.
\end{equation}
	
\section{Ray tracing and geodesic deviation equations}
\label{sec:Null+geod+Kerr}
\subsection{Null geodesics and the null geodesic deviation equation in Kerr spacetime}

In the simulation of the lensing effects of the material surrounding the immediate vicinity of 
Sgr~A$^*$ we will assume as in \cite{Boero:2021} that the underling geometry is
given by the two parametric $(M,a)$ Kerr line element:
\begin{equation}\label{eq:Kerr-usualBL-form}
\begin{split}
ds^2 
=& \left( 1 - \varPhi\right) dt^2  
+ 2 \varPhi a \sin(\theta)^2 dt d\phi 
- \frac{\Sigma}{\Delta} dr^2 
\\
& - \Sigma d\theta^2 
- \left( r^2 + a^2 + \varPhi a^2  \sin^2(\theta)\right) \sin(\theta)^2
d\phi^2.
\end{split}
\end{equation}
As usual $M$ denotes the mass of the spacetime and $a$ the rotation parameter respectively and 
the functions $\Sigma(r,\theta)$, $\Delta(r)$ and $\varPhi(r,\theta)$ 
are given by
\begin{equation}\label{eq:metric+func+Sigma}
\Sigma = r^2 + a^2 \cos(\theta)^2 , 
\end{equation}
\begin{equation}\label{eq:eq:metric+func+Delta}
\Delta = r^2 - 2rM + a^2 ,  
\end{equation}
\begin{equation}\label{eq:eq:metric+func+Phi}
\varPhi =\frac{2 M r}{\Sigma}.
\end{equation}

Beyond being quite ubiquitous to describe the expected gravitational field in the neighbourhood 
of the central core of a galaxy, Kerr metric has the nice feature that both geodesics and null 
geodesic deviation equations acquire expressions that permit a relatively easy treatment \cite{Carter:1966zza, Boero:2019zkq}.   

For numerical purposes the full system of equations can be cast as a first order system as follows:
\begin{align}
\dot t 
=& 
\frac{1}{\Sigma \Delta}\left[ E \Big(
(r^2 + a^2)^2 - \Delta \, a^2 \sin(\theta)^2 \Big)
- 2 a M r L_z \right]
, 
\\
\dot{r} =& v^r,
\\
\dot{v}^r =&  
(v^{\theta})^2 r \frac{\Delta}{\Sigma} 
+ \frac{a^2}{\Sigma} \dot{r} \dot{\theta} \sin(2\theta) 
- (v^{r})^2 \frac{r \Delta + \left(M - r \right)\Sigma}{\Sigma \Delta} \nonumber 
\\
& 
- \dot{t}^2 
\frac{ M \Delta \big( r^2 - a^2 \cos(\theta)^2 \big)}{\Sigma^3} \nonumber
\\
& + 2 \dot{t}\dot{\phi}
\frac{ a M \Delta \big( r^2 - a^2 \cos(\theta)^2 \big) \sin(\theta)^2}{\Sigma^3} \nonumber
\\
&
+
\dot{\phi}^2 
\frac{\Delta \sin(\theta)^2}{\Sigma^3}
\Big( r\Sigma^2 \nonumber \\
& \qquad \qquad \qquad \;\; - a^2 M\big( r^2 - a^2\cos(\theta)^2\big)\sin(\theta)^2 \Big)
,
\\
\dot{\theta} =& v^\theta, 
\\
\dot{v}^\theta =& \frac{a^2 M r \sin(2\theta)}{\Sigma^3} \dot{t}^2 
- \frac{2aMr \left( r^2 + a^2 \right)\sin(2\theta)}{\Sigma^3}\dot{t}\dot{\phi} \nonumber 
\\
& - \frac{a^2 \sin(2\theta)}{2 \Sigma \Delta} (v^{r})^2 
- \frac{2r}{\Sigma}\dot{r}\dot{\theta} +  
\frac{a^2 \sin(2\theta)}{2 \Sigma} (v^{\theta})^2  \nonumber
\\
&+ \frac{\sin(2\theta)}{2\Sigma^3} 
\Big( \left( r^2 + a^2\right)\Sigma^2 
\nonumber
\\
& \qquad \qquad \quad + 2a^2rM\sin(\theta)^2 
\left( r^2 + a^2 + \Sigma \right) \Big) \dot{\phi}^2
,
\\
\dot \phi =& \frac{1}{\Sigma\Delta}\left[ 2E a M r + (\Sigma - 2Mr) \frac{L_z}{\sin(\theta)^2}\right]
,
\end{align}
\begin{align}
\dot{\varsigma}_{R1}^{} =& \upsilon^{\varsigma}_{R1}, 
\\
\dot{\upsilon}^{\varsigma}_{R1} =& - \varsigma_{R1}^{} \Psi_{0R}^{} - \varsigma_{I1}^{} \Psi_{0I}^{},
\\
\dot{\varsigma}_{I1}^{} =& \upsilon^{\varsigma}_{I1},
\\
\dot{\upsilon}^{\varsigma}_{I1} =& - \varsigma_{R1}^{} \Psi_{0I}^{} + \varsigma_{I1}^{} \Psi_{0R}^{},
\end{align}
\begin{align}
\dot{\varsigma}_{R2}^{} =& \upsilon^{\varsigma}_{R2}, 
\\
\dot{\upsilon}^{\varsigma}_{R2} =& - \varsigma_{R2}^{} \Psi_{0R}^{} - \varsigma_{I2}^{} \Psi_{0I}^{},
\\
\dot{\varsigma}_{I2}^{} =& \upsilon^{\varsigma}_{I2},
\\
\dot{\upsilon}^{\varsigma}_{I2} =& - \varsigma_{R2}^{} \Psi_{0I}^{} + \varsigma_{I2}^{} \Psi_{0R}^{}.
\end{align}  
In the above expressions, the constants of motion $E$ and $L$ along the central null geodesics 
are obtained through its relation to the coordinates $(r_o, \theta_o)$ of an stationary observer and the directions of the 
incoming photons $ \left(\alpha_x, \delta_z \right) $ through the following expressions:
\begin{equation}
E = -\sqrt{1 - \varPhi_o},
\end{equation}
\begin{align}
\alpha_{x}
=&
\frac{L_z}{r_o \sin(\theta_o)}
, 
\\
\begin{split}
\delta_{z} 
=&
-
\frac{(\pm)}{r_o} 
\left[K - \left( \frac{L_z}{\sin(\theta_o)} 
- aE\sin(\theta_o) \right)^2\right]^{1/2}
,
\end{split}
\end{align}
where $K$ is Carter's constant. It should be noticed that in the numerical calculations we first
choose $\alpha_{x}$ and $\delta_{z}$, from which
we infer, at the observer position, the values
of the constant of motion $L_z$ and $K$;
although $K$ does not appear in the evolution
equations; but they both appear in the initial
conditions.
The Weyl curvature scalar $\Psi_0 = \Psi_{0R} + i \Psi_{0I}$
is given by the compact expression\citep{Boero:2019zkq}: 
\begin{equation}\label{eq:Psi0}
\Psi_0 = - \frac{3 M^{5/3} \mathbb{K}^2}{2 \big(r - ia \cos(\theta) \big)^5}; 
\end{equation}
where the constant $\mathbb{K}$ is a spin-weight quantity given by:
\begin{equation}\label{eq:K}
\begin{split}
\mathbb{K} =& 
\frac{i \sqrt{2}}{ M^{1/3}}\left[
\delta_z r_o
- 
i\big(- aE\sin(\theta_o)  
+ \alpha_x r_o \big)\right],
\end{split}
\end{equation}
and satisfies $\mathbb{K} \bar{\mathbb{K}} = 2 M^{-2/3} K^2$. 

The appropriated initial conditions for the geodesic equation are
\begin{align}
t_0 =& t_o = 0, \\
r_0 =& r_o, \\
v^r_0 =& - \frac{\sqrt{\mathcal{R}(r_o)}}{\Sigma_o} , \\
\theta_0 =& \theta_o, \\
\ell^\theta_0 =& \pm \frac{\sqrt{\varTheta(\theta_o)}}{\Sigma_o}, \\
\phi_0 =& \phi_o = - \frac{\pi}{2}
,
\end{align}
where the functions $\mathcal{R}(r)$ and $\varTheta(\theta)$ are defined as
\begin{align}
\mathcal{R}(r) =& \Big( E\left(r^2 + a^2 \right) - a L \Big)^2 - K \Delta(r) 
,
\label{eq:R+geod+func}
\\
\varTheta(\theta) =& K - \left( \frac{L}{\sin(\theta)} - a E \sin(\theta) \right)^2 
.
\end{align}
For the geodesics deviation equations one can use for example either of the choices previously described in the section \ref{sec:Exact+grav+lens+opt+scalars}, namely
the set of initial conditions \eqref{eq:IC_firstchoie_devgeod_0} - \eqref{eq:IC_firstchoie_devgeod_3} or
\eqref{eq:IC_secondchoie_devgeod_0} - \eqref{eq:IC_secondchoie_devgeod_3}.

The above system of equations constitute the base of our ray-tracing 
method which efficiently includes magnification effects.
We used a code written in FORTRAN90 working in quadruple precision,
and making use of a 7–8 Runge–Kutta solver contained in the RKSUITE\citep{rksuite-90}
library.
As noted previously in reference \cite{Boero:2021}, it achieves great accuracy as revealed by the 
very small relative errors 
of the conserved quantities 
along the null paths.
For other treatments of ray-tracing see for example
\cite{Vincent:2011wz, Bronzwaer:2018lde, M:2022pex, Chan:2017igo, Pu:2016eml, Pelle:2022phf}.
We also point out that photon conservation is implicit in our ray-tracing. The discussion of radiative transfer effects as well as the
treatment of polarization will be addressed in a future works along the lines presented early in \cite{Boero:2019zkq}. 
These are features discussed in numerical codes like
\cite{ Dexter:2016cdk, Pihajoki:2018ihj, Bronzwaer:2020kle}
that we would like to include in future work.

\subsection{Simple accretion disc model in Kerr}\label{subsec:Simple_AccDisc}

The working assumption in this work 
is of a geometrically thin disk,
where the matter motion is confined to the equatorial 
plane, with circular orbits.
Our approximation to this system is mainly supported
from our previous work on M87, where a similar
model gave excellent results. However we are
aware that this is a model with limitations;
as already mentioned in \cite{Paczynsky:1979rz}.
Nevertheless thin disk models have remain as one
of the persisting models of accretion 
disks\citep{Shakura1973,Abramowicz:2011xu}.
It should also be emphasized that most of the
observations on the nature of 
Sgr~A$^*$\citep{Baganoff:2001ju,2019Natur.570...83M}
had been carried out through measurements 
involving a single observatory, and therefore their
work and description involves dimensions
which are about four orders of magnitude
bigger than the region observed and studied by the
EHT team.
When studying a complicated system it is customary
to start with the simplest model and then continue
with more complex and descriptive ones.
So, our choice for a thin-disk model, must be
understood in this framework. We do plan to explore
in the  future other more complicated structures
as 3D geometries and ambient plasma, involving
more detailed models of emission.
This would allow to compare more realistic features present in simulations\citep{Font:2008fka} like possible non-axisymmetric structures, turbulence 
and the full influence of absorption in the radiative transfer.

Assigning the label 'e' to an emitter, then its
motion is characterized by the values of radial
and angular coordinates $(r=r_e,\theta=\pi/2)$,
and by its four velocity 
$u_e^a = \left( \dot{t}_e, 0, 0, \dot{\phi}_e \right)^a$;
in such a way that we have the following
equations of motion:
\begin{equation}\label{eq:tdot}
\begin{split}
r_e^2 \dot t_e 
=& 
\frac{1}{\Delta_e}\left[
E_e \bigg(
(r_e^2 + a^2)^2 - \Delta \, a^2 
\bigg)
- 2 a M r_e L_e\right]
,
\end{split}
\end{equation}
\begin{equation}\label{eq:rdot}
\begin{split}
0 
=& \Big( E_e (r_e^2 + a^2) - a L_e \Big)^2 - \Delta (r_e^2 + K_e)
,
\end{split}
\end{equation}
\begin{equation}\label{eq:titadot}
\begin{split} 
0 
=& K_e - (E_e a - L_e)^2
,
\end{split}
\end{equation}
\begin{equation}\label{eq:fidot}
\begin{split}
r_e^2 \dot \phi_e
=& \frac{1}{\Delta}\Big[ 2 E_e a M r_e + (r_e^2 - 2 M r_e) L_e \Big]
.
\end{split}
\end{equation}

As explained in our previous article, the constants
of motion can be calculated, following 
\cite{Chandrasekhar:1983este}, so that, using
\begin{equation}
u=1/r_e,
\end{equation}
and 
\begin{equation}\label{key}
Q_\pm = 1 - 3M u \pm 2 a \sqrt{M u^3}
;
\end{equation}
one then obtains that the values of the energy $E_e$ and angular momentum $L_e$ 
are given by:
\begin{equation}\label{key}
E_e = 
\frac{1}{\sqrt{Q_\mp}}
\Big(
1 - 2 M u \mp a \sqrt{M u^3}
\Big) 
,
\end{equation}
and
\begin{equation}\label{key}
L_e =
\mp
\frac{\sqrt{M}}{\sqrt{u Q_\mp}}
\Big(
a^2 u^2 + 1 \pm 2 a\sqrt{M u^3}
\Big)
;
\end{equation}
where upper sing applies to retrograde orbits while lower sign applies to direct orbits.

The range of validity of the these equations
has been discussed in our previous article;
we here just note that calling $r_c$ the
relevant root of $Q$, then for $r_+ \le r < r_c$
we will use the same values of energy and
angular momentum, as those for the last
stable orbit.

In this respect,
let us recall that the unstable circular photon orbit
on the equatorial plane is given by
\begin{equation}\label{key}
r_c = 2 M \Bigg(
1 + \cos\left(
\frac{2}{3} \arccos\left(\pm \frac{a}{M}\right)
\right)
\Bigg)
;
\end{equation}
where upper sing applies to retrograde orbits while lower sign applies to direct orbits.

For the calculation of the flux we need the factor
\begin{equation}
1/(1+z)^4 = 1/(u_e^b \ell_b)^4,   
\end{equation}
where
\begin{equation}\label{eq:ue_ell}
\begin{split}
g_{ab} u_e^a \ell^b 
=&
E_e \dot t - L_e \dot \phi 
.
\end{split}
\end{equation}

\section{
Simulated images with gravitational lens magnification and red/blueshift effects
}\label{sec:Results}

\subsection{Basic model}
In this work we use for the mass of the supermassive
black hole the value published in 
\cite{GRAVITY:2021xju}, which is
$M=(4.297 \pm 0.012) \times 10^6 \text{M}_{\astrosun}$.
For the distance to Sgr~A${}^*$ we use the 
value of $8.178\pm 0.035$kpc from
\cite{Gravity:2019AyA}.

The direction of the angular momentum was deduced from reference \cite{Reid2020},
interpreting the apparent motion of Sgr~A${}^*$, as due to the motion of the solar
system, and its surroundings around the center supermassive black hole.
In this way we assign the direction of the angular momentum to be
$30.22^\circ$ south of west direction.

Beside the geometric configuration an emission profile must be given as input in 
order to predict the observed images.
We choose an emission model having two temperature regions \cite{Moscibrodzka:2015pda} in a similar way as we used in our previous
article.
As explained in their article this model captures the main 
characteristics of the equatorial emission in GRMHD simulations. 

\subsection{Image from the EHT Collaboration of Sagittarius A${}^*$}

In Fig. \ref{fig:eht-image} we reproduce the EHT image of Sgr~A${}^*$, 
that we use as our main reference for this work.
The image is a reconstruction made from observations taken on April 7 of 2017; 
it consists of the average of another four images, each one of them obtained from a larger 
set of reconstructions.
According to the EHT, these four groups were a convenient way to manage the analysis 
of the observed differences shown by the several resulting images with the diverse processing 
pipelines employed. 
Three of the four clusters were intended to separate ring structures having a salient 
bright spot at different sectors of the image and the fourth one group intended to detect 
non-ring morphologies.
The relative contribution of each cluster to the final image can be seen in fig. 13 of 
\cite{EventHorizonTelescope:2022wok}; in particular, it shows that most of the reconstruction 
obtained are those having bright spots at north-west and south-west position angle (PA).
Then, the results presented exhibit some dependency on the choice of parameter used by the 
imaging pipelines as well as the imagine procedure itself.
The observational challenges related with the imaging process of Sgr~A${}^*$ and the 
remaining degeneracy in the final results leave several open questions.
For instance, does it becomes appropriated to consider averages of a set of images in order 
to picture features of the object under study?
Since different reconstruction methods produce different features one could ask
which of them are physically representative of the source.
A common feature that also appears in most of the reconstructed images is a bright ring,
but since the reconstruction algorithm needed additional input, the question still remains 
if the choices of parameter has been the
optimum.
Some of these points have been addressed in reference \cite{EventHorizonTelescope:2022wok} 
(see section 7.5) pointing that non-ring structures are very unlikely.

The presence of an elliptical ring structure
in the image of EHT, 
suggests a source which is observed 
fairly face-on. This has been our first assumption
that we study in detail in subsection \ref{subsec:angellip}: while in subsection
\ref{subsec:ang-galax} we will consider another
geometry.

Beyond the three brighter regions shown in the image; it is somehow striking the elliptical shape, 
of the visible background structure, that it could suggest a matter distribution more or less with 
the shape of an inclined disk.
What is striking is that the angle of the semimajor axis, of this structure,
is not contained in the plane of the galaxy.
Also the bright region on the top of this image, supports this interpretation of an inclined disk.

In any case, for these reasons we have studied the possibility that there were a thin disk with 
the orientation suggested for this image. This is analysed in the next subsection \ref{subsec:angellip}.
But also, in subsection \ref{subsec:ang-galax} we study the more natural expected
situation, that the projected angular momentum of the black holeis aligned with the
angular momentum of the galaxy.


\subsection{Graphs with angular momentum aligned 
perpendicularly to the elliptical
shape of the EHT image}\label{subsec:angellip}

In this subsection we show the graphs corresponding to images
calculated with an angular momentum of the black hole, whose projection coincide with
the perpendicular direction to the semi-mayor axis of the elliptical 
image published by the EHT Collaboration team for Sgr~A${}^*$.
We use the following values of iota:
$\iota = -0.1745$, 
$\iota = -0.3490$, 
$\iota = -0.5235$, 
$\iota = -0.6980$, 
$\iota = -0.8725$, 
$\iota = -1.0470$, 
$\iota = -1.2215$, 
$\iota = -1.3960$;
corresponding to degrees of
$-10^o$, 
$-20^o$, 
$-30^o$, 
$-40^o$, 
$-50^o$, 
$-60^o$, 
$-70^o$, 
$-80^o$, 
of the angle of the black hole angular momentum with the plane of the image.
From left to right, in each figure we show the images corresponding to 
$a=0.98$,  $a=0.75$, $a=0.50$, $a=0.25$ and $a=0.00$ respectively.

The comparison of images is very
complicated, and it is difficult
to have a universal way to be
applied to any situation. In our
case, we use our understanding
of the possible astrophysical situation, but we also use
a direct numerical measure employing
the correlation coefficient
of the comparison of the EHT image
with our images. This is not
an ideal measure, but it provides
a minimum information of coincidence
between images.
In particular we will see in this
set that high values of the
correlation coefficient do not
necessarily provide good
coincidence of the images.
We have encounter the same problematic
in our previous work \citep{Boero:2021}
which can be consulted for further
explanation.

In Fig. \ref{fig:rho-b} we present
the graph of the correlation coefficient between 
the EHT image and our images.

\begin{figure*}
\centering
\includegraphics[clip,width=0.195\textwidth]{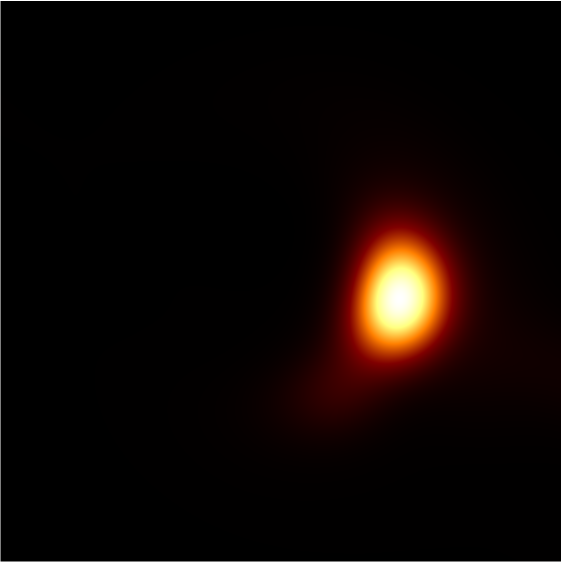}
\includegraphics[clip,width=0.195\textwidth]{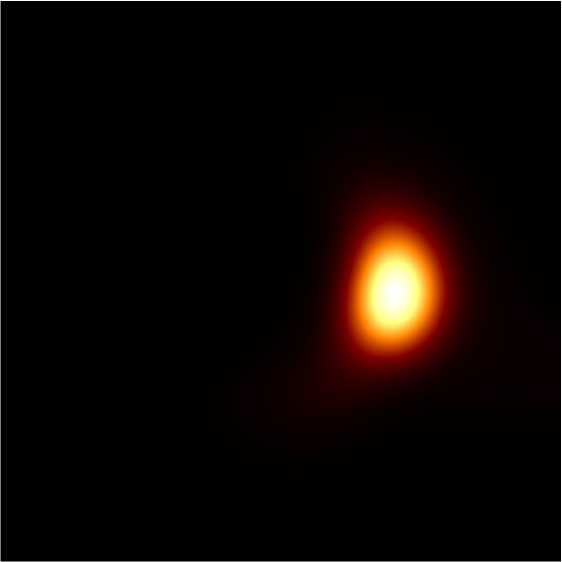}
\includegraphics[clip,width=0.195\textwidth]{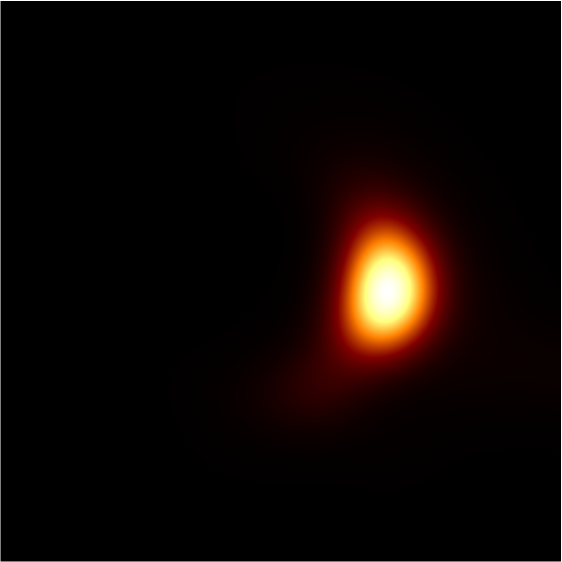}
\includegraphics[clip,width=0.195\textwidth]{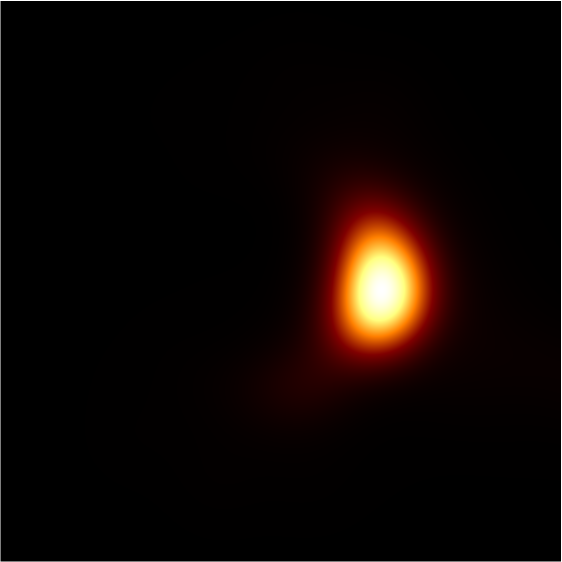}
\includegraphics[clip,width=0.195\textwidth]{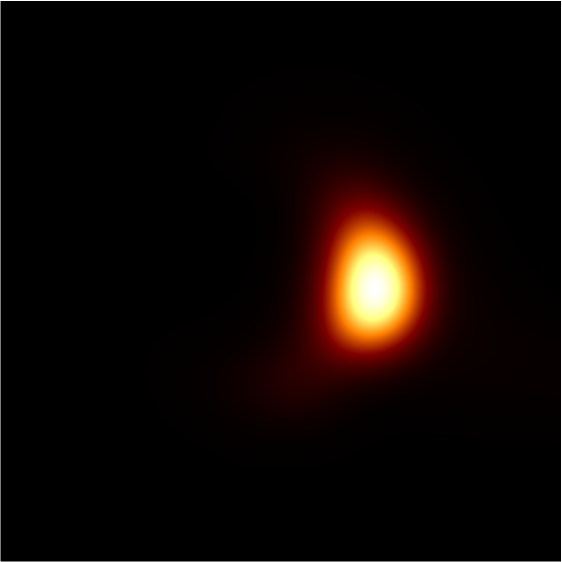}
\caption{$\iota=-0.1745=-10^\circ$, and from left to right with Kerr parameter: 
	$a=0.00$,  $a=0.25$, $a=0.50$, $a=0.75$ and $a=0.98$.
}
\label{fig:io-1745}
\end{figure*}

\begin{figure*}
\centering
\includegraphics[clip,width=0.195\textwidth]{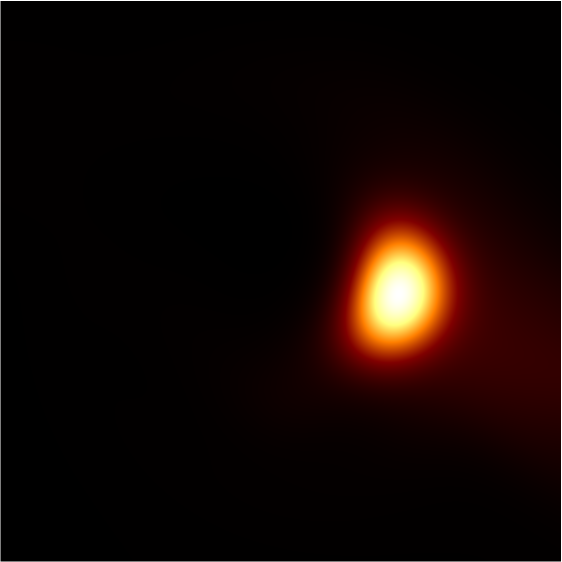}
\includegraphics[clip,width=0.195\textwidth]{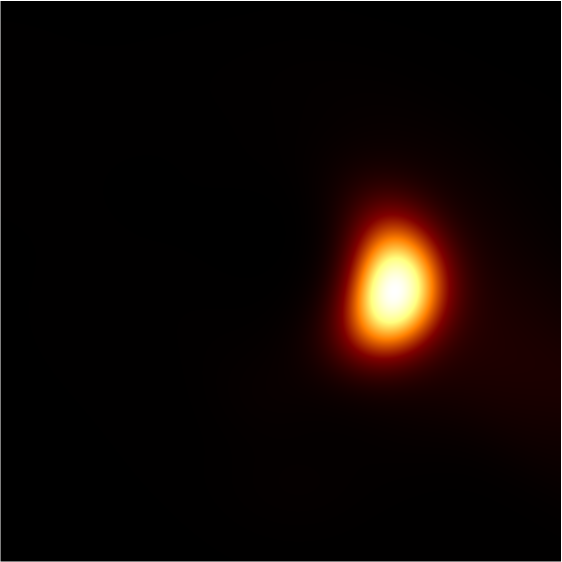}
\includegraphics[clip,width=0.195\textwidth]{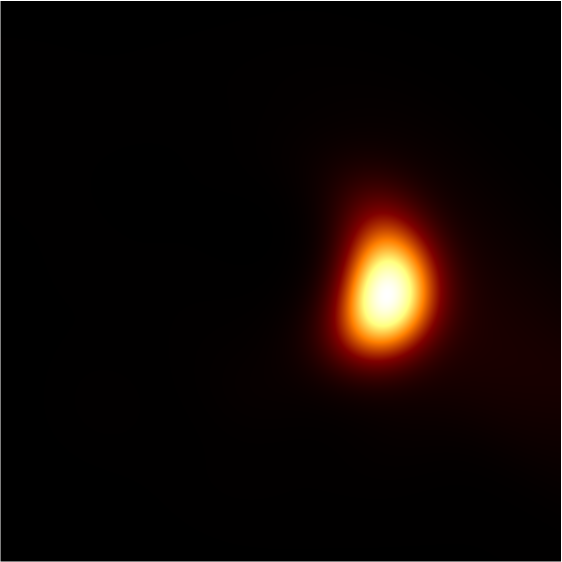}
\includegraphics[clip,width=0.195\textwidth]{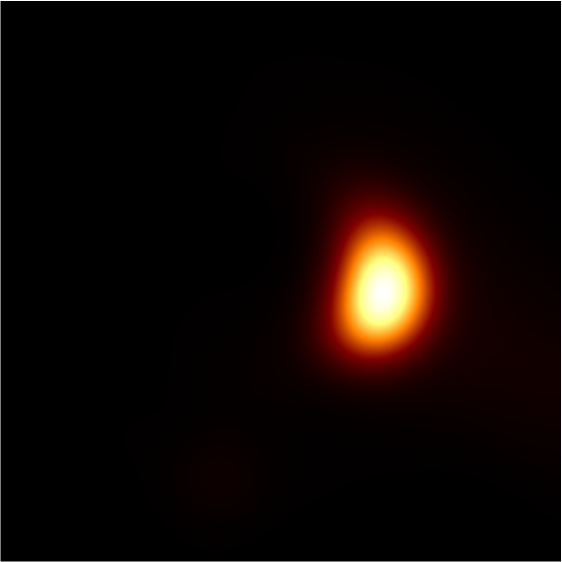}
\includegraphics[clip,width=0.195\textwidth]{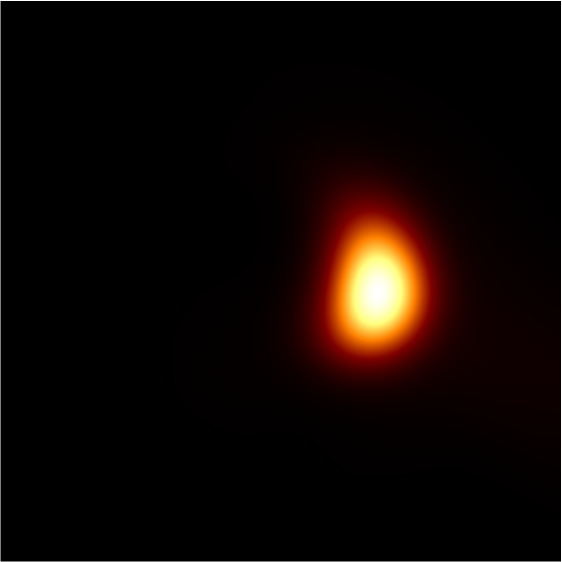}
\caption{$\iota=-0.3490=-20^\circ$, and from left to right with Kerr parameter: 
	$a=0.00$,  $a=0.25$, $a=0.50$, $a=0.75$ and $a=0.98$.
}
\label{fig:io-3490}
\end{figure*}

\begin{figure*}
\centering
\includegraphics[clip,width=0.195\textwidth]{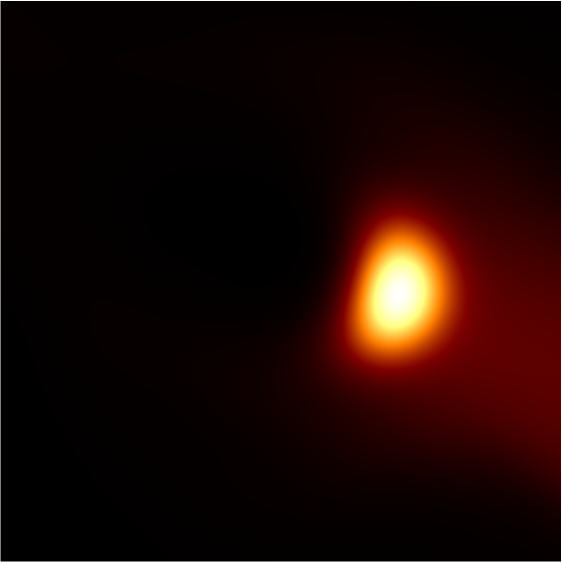}
\includegraphics[clip,width=0.195\textwidth]{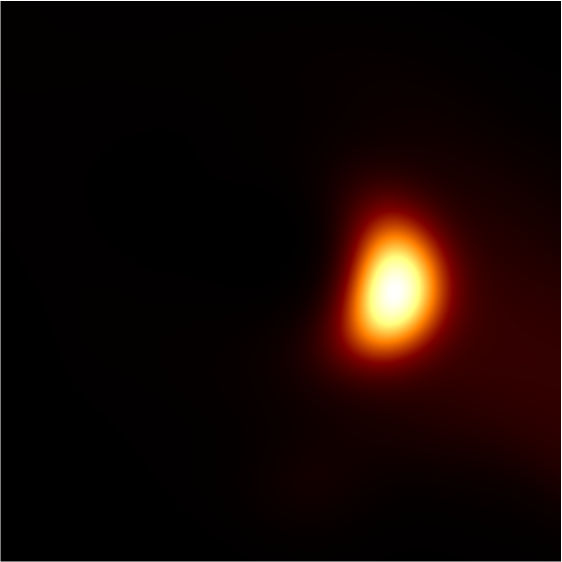}
\includegraphics[clip,width=0.195\textwidth]{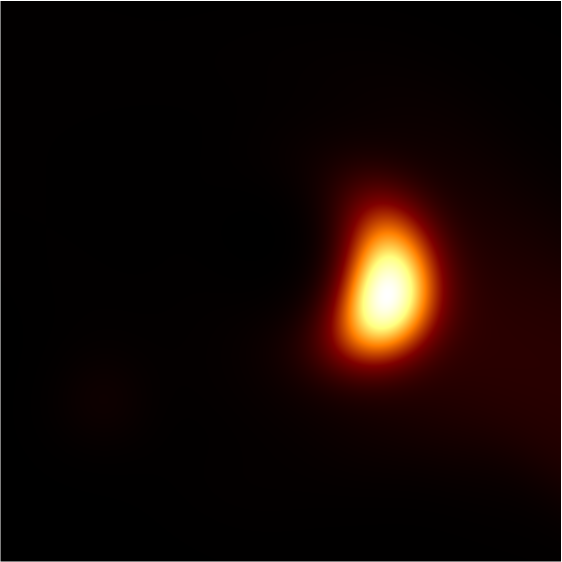}
\includegraphics[clip,width=0.195\textwidth]{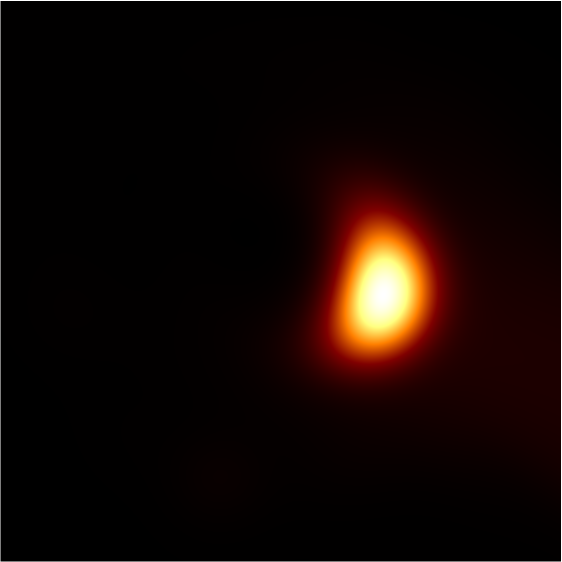}
\includegraphics[clip,width=0.195\textwidth]{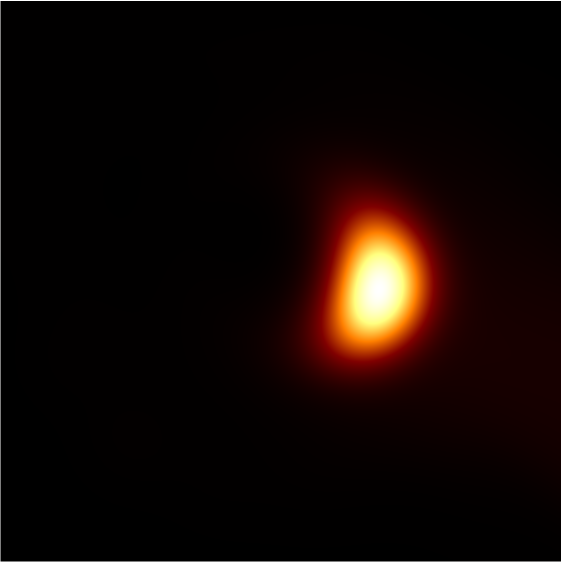}
\caption{$\iota=-0.5235=-30^\circ$, and from left to right with Kerr parameter:
		$a=0.00$,  $a=0.25$, $a=0.50$, $a=0.75$ and $a=0.98$.
}
\label{fig:io-5235}
\end{figure*}

\begin{figure*}
\centering
\includegraphics[clip,width=0.195\textwidth]{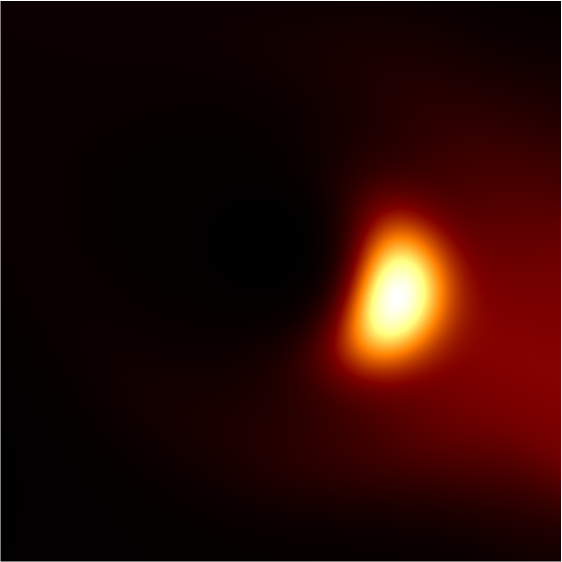}
\includegraphics[clip,width=0.195\textwidth]{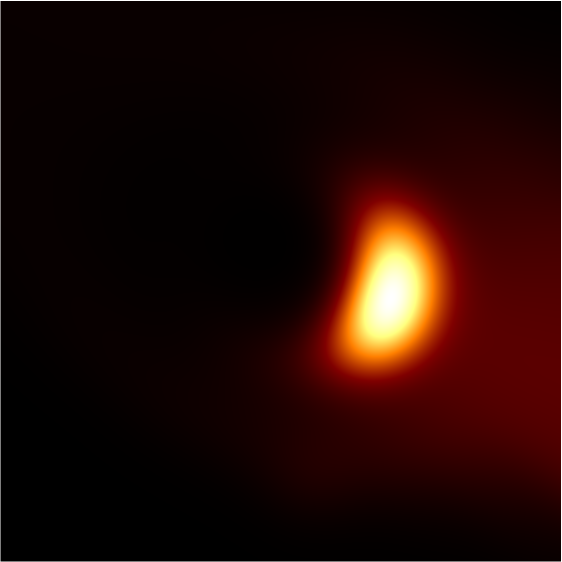}
\includegraphics[clip,width=0.195\textwidth]{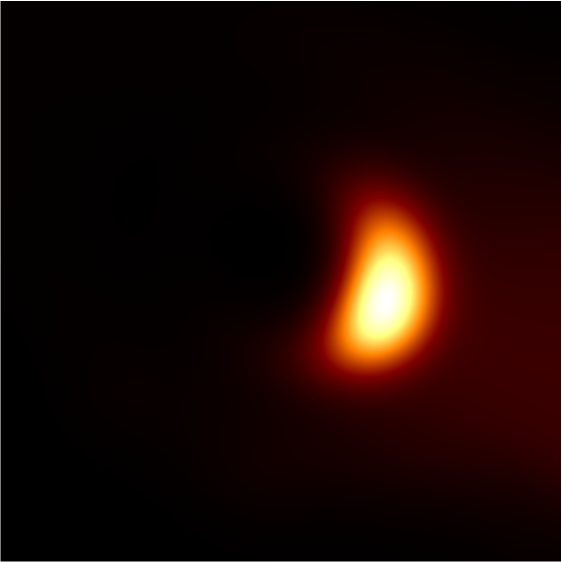}
\includegraphics[clip,width=0.195\textwidth]{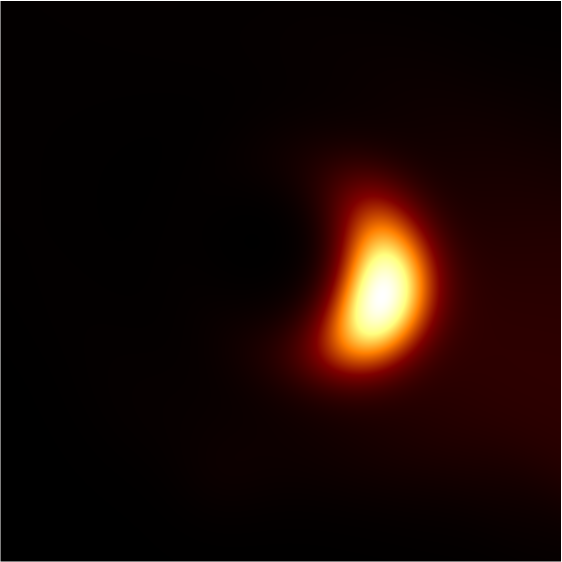}
\includegraphics[clip,width=0.195\textwidth]{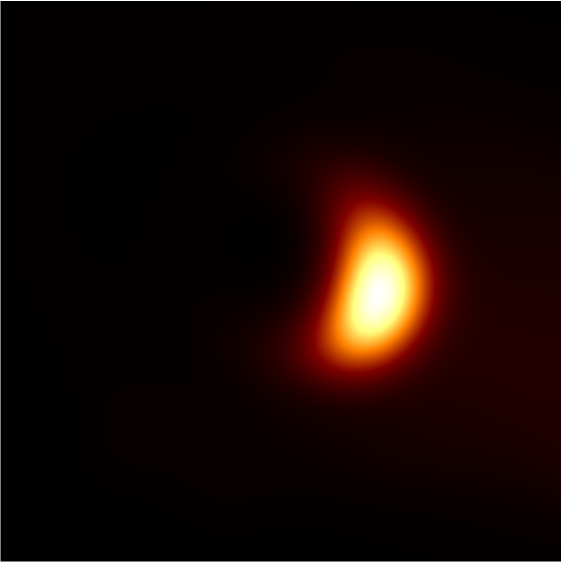}
\caption{$\iota=-0.6980=-40^\circ$, and from left to right with Kerr parameter: 
		$a=0.00$,  $a=0.25$, $a=0.50$, $a=0.75$ and $a=0.98$.
}
\label{fig:io-6980}
\end{figure*}

\begin{figure*}
\centering
\includegraphics[clip,width=0.195\textwidth]{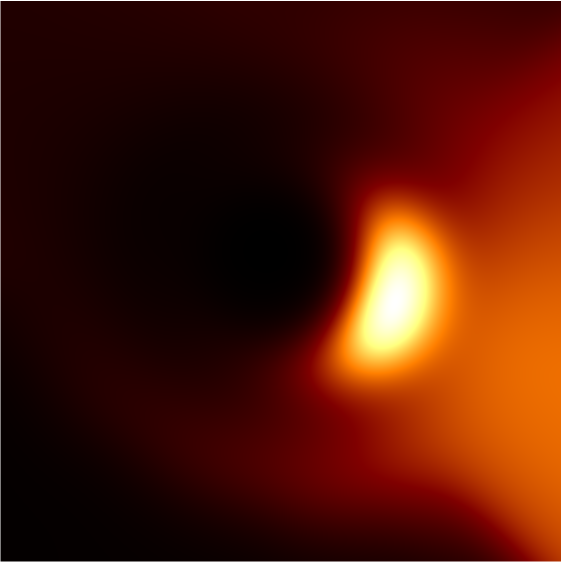}
\includegraphics[clip,width=0.195\textwidth]{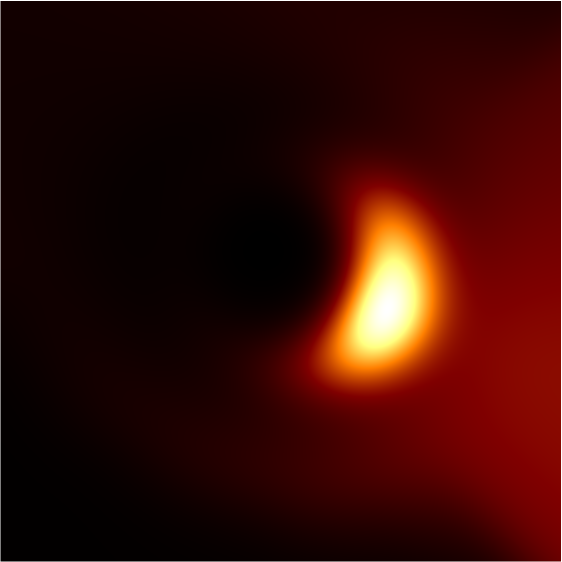}
\includegraphics[clip,width=0.195\textwidth]{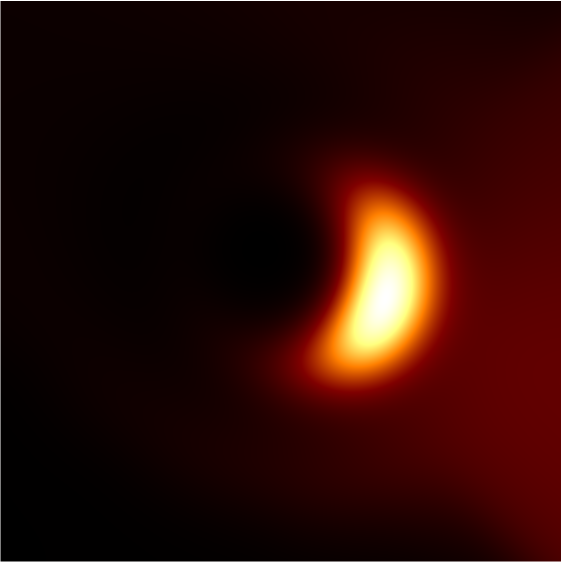}
\includegraphics[clip,width=0.195\textwidth]{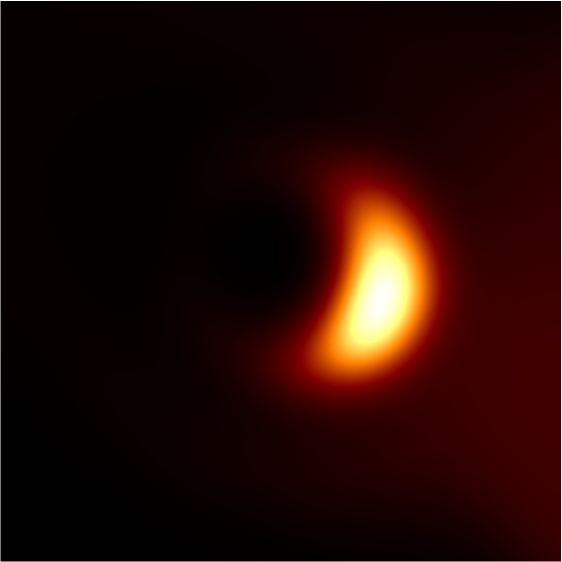}
\includegraphics[clip,width=0.195\textwidth]{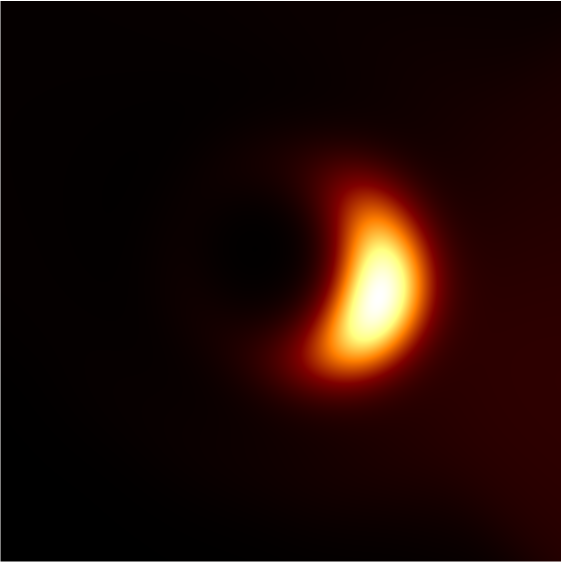}
\caption{$\iota=-0.8725=-50^\circ$, and from left to right with Kerr parameter: 
		$a=0.00$,  $a=0.25$, $a=0.50$, $a=0.75$ and $a=0.98$.
}
\label{fig:io-8725}
\end{figure*}

\begin{figure*}
\centering
\includegraphics[clip,width=0.195\textwidth]{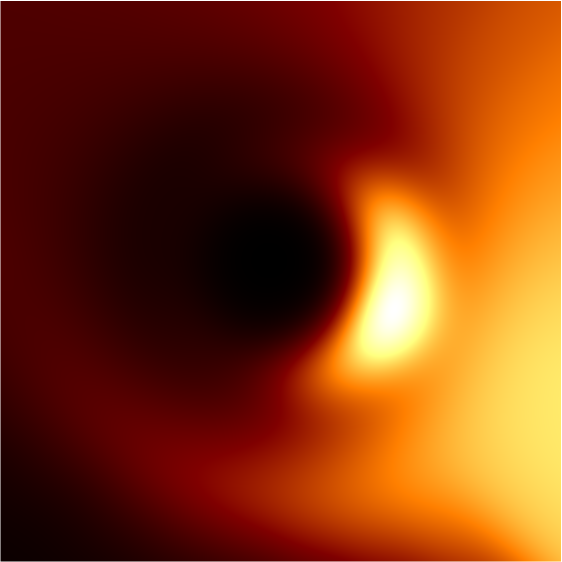}
\includegraphics[clip,width=0.195\textwidth]{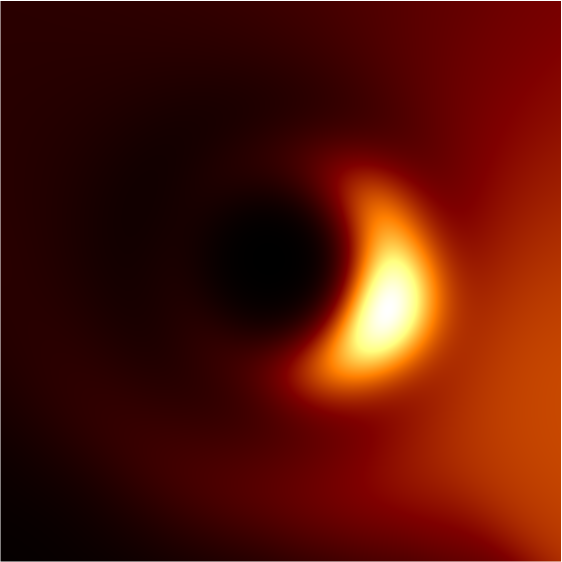}
\includegraphics[clip,width=0.195\textwidth]{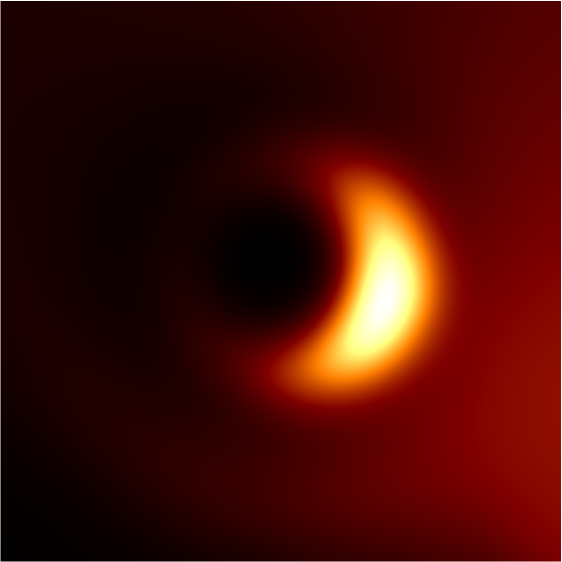}
\includegraphics[clip,width=0.195\textwidth]{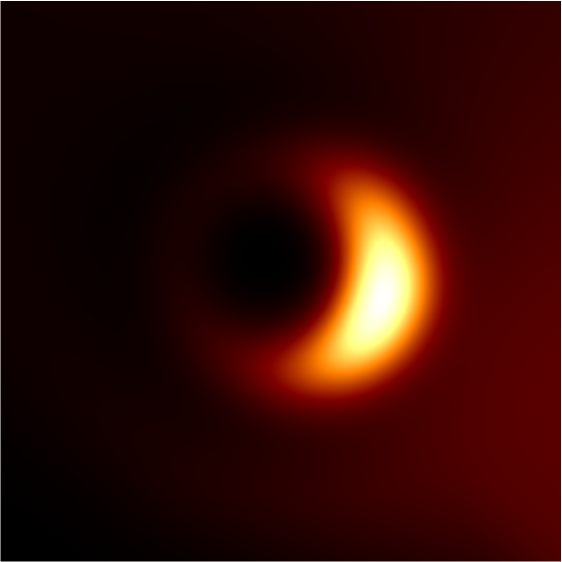}
\includegraphics[clip,width=0.195\textwidth]{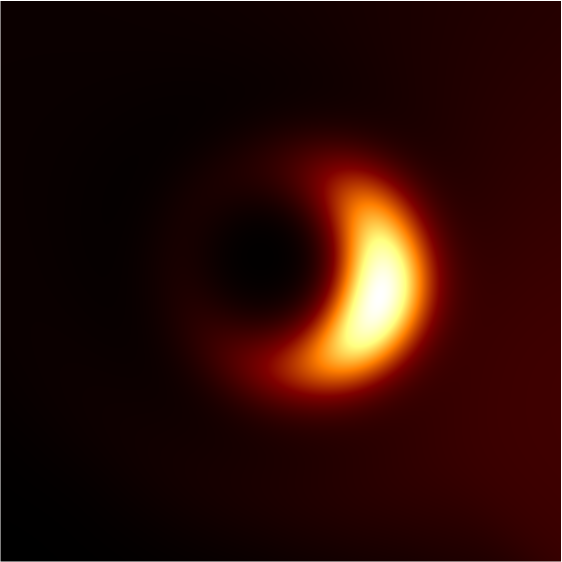}
\caption{$\iota=-1.0470=-60^\circ$, and from left to right with Kerr parameter: 
		$a=0.00$,  $a=0.25$, $a=0.50$, $a=0.75$ and $a=0.98$.
}
\label{fig:io-1,0470}
\end{figure*}

\begin{figure*}
\centering
\includegraphics[clip,width=0.195\textwidth]{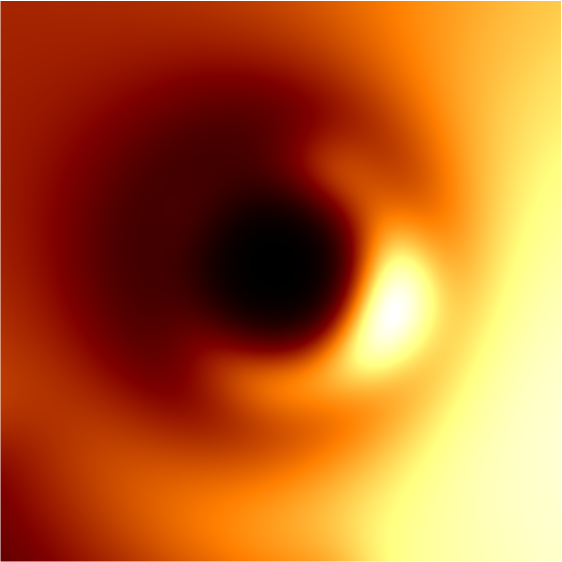}
\includegraphics[clip,width=0.195\textwidth]{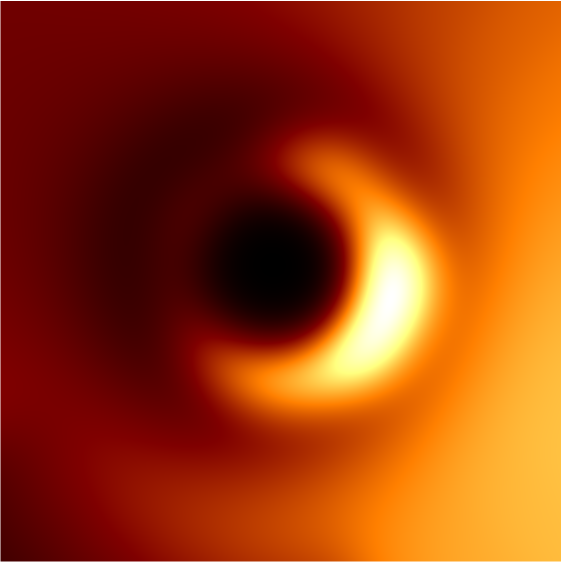}
\includegraphics[clip,width=0.195\textwidth]{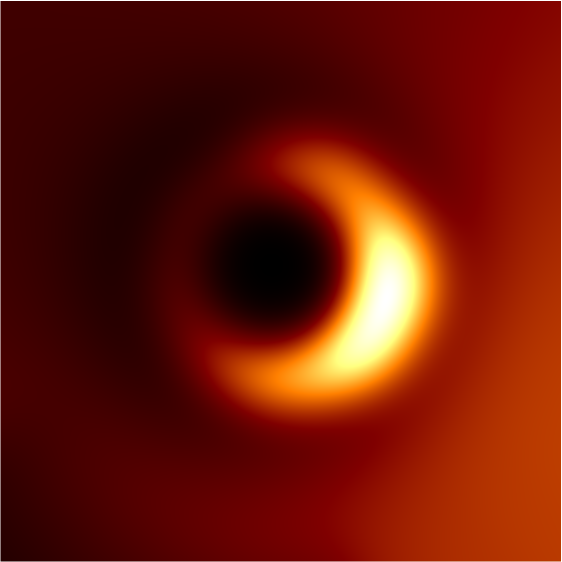}
\includegraphics[clip,width=0.195\textwidth]{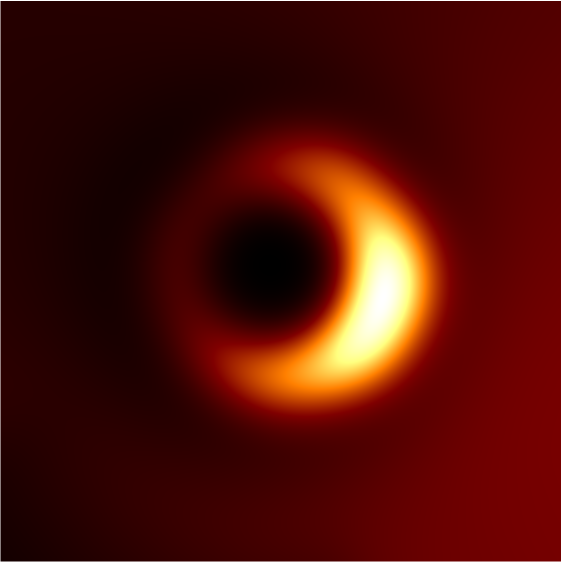}
\includegraphics[clip,width=0.195\textwidth]{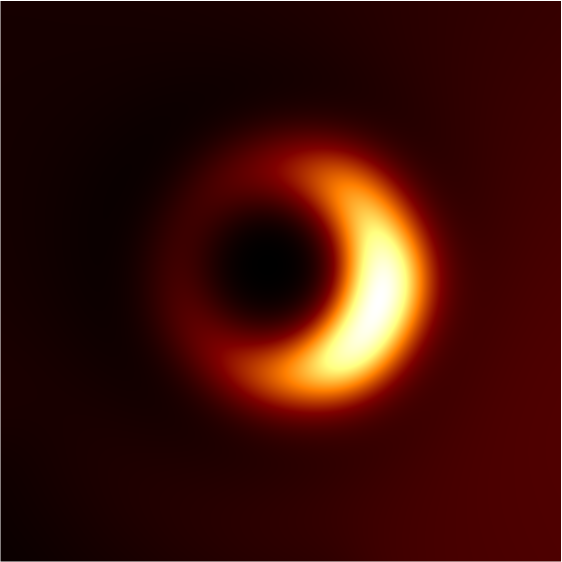}
\caption{$\iota=-1.2215=-70^\circ$, and from left to right with Kerr parameter: 
		$a=0.00$,  $a=0.25$, $a=0.50$, $a=0.75$ and $a=0.98$.
}
\label{fig:io-1,2215}
\end{figure*}

\begin{figure*}
\centering
\includegraphics[clip,width=0.195\textwidth]{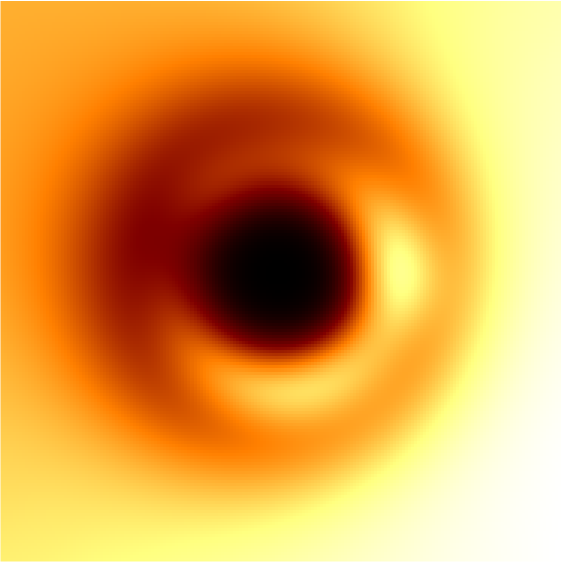}
\includegraphics[clip,width=0.195\textwidth]{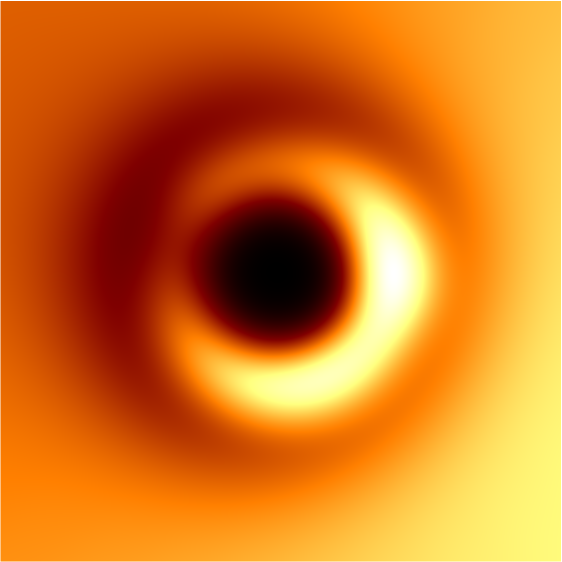}
\includegraphics[clip,width=0.195\textwidth]{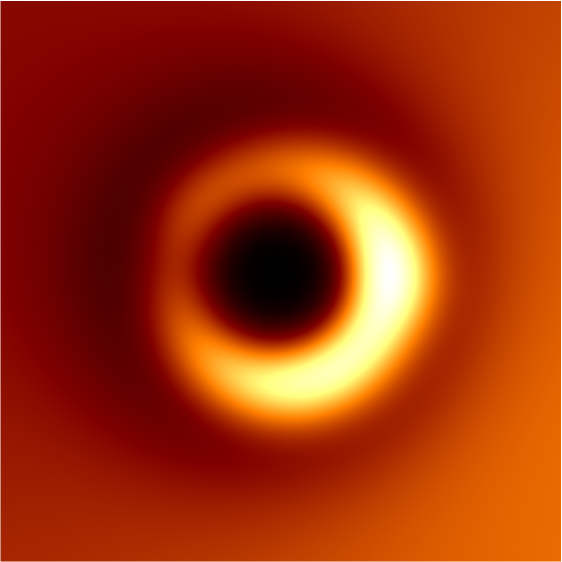}
\includegraphics[clip,width=0.195\textwidth]{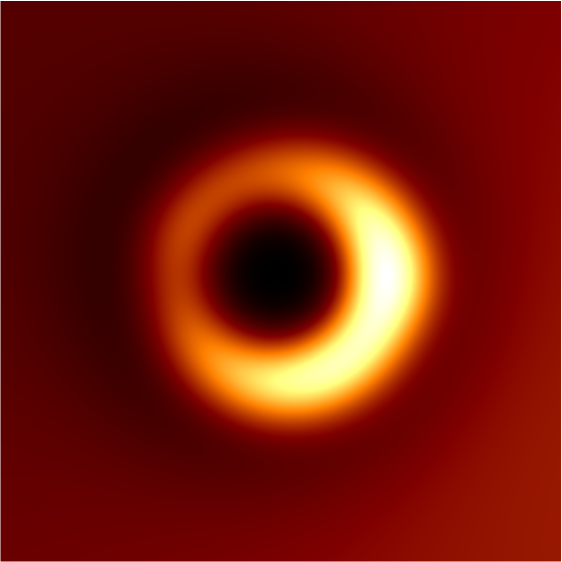}
\includegraphics[clip,width=0.195\textwidth]{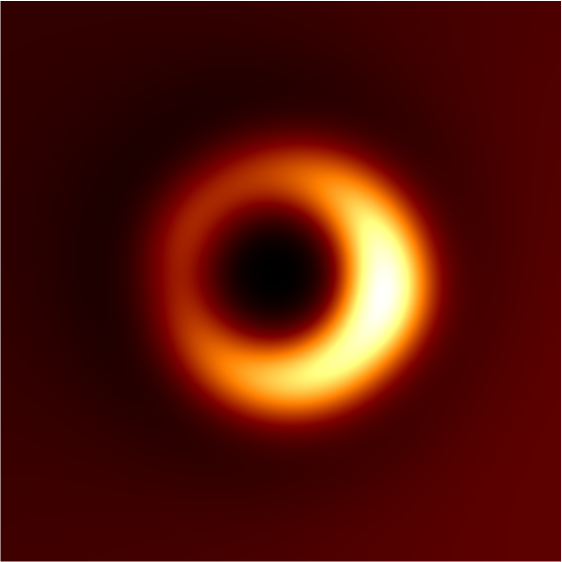}
\caption{$\iota=-1.3960=-80^\circ$, and from left to right with Kerr parameter: 
		$a=0.00$,  $a=0.25$, $a=0.50$, $a=0.75$ and $a=0.98$.
}
\label{fig:io-1,3960}
\end{figure*}

\begin{figure*}
\centering
\includegraphics[clip,width=0.7\textwidth]{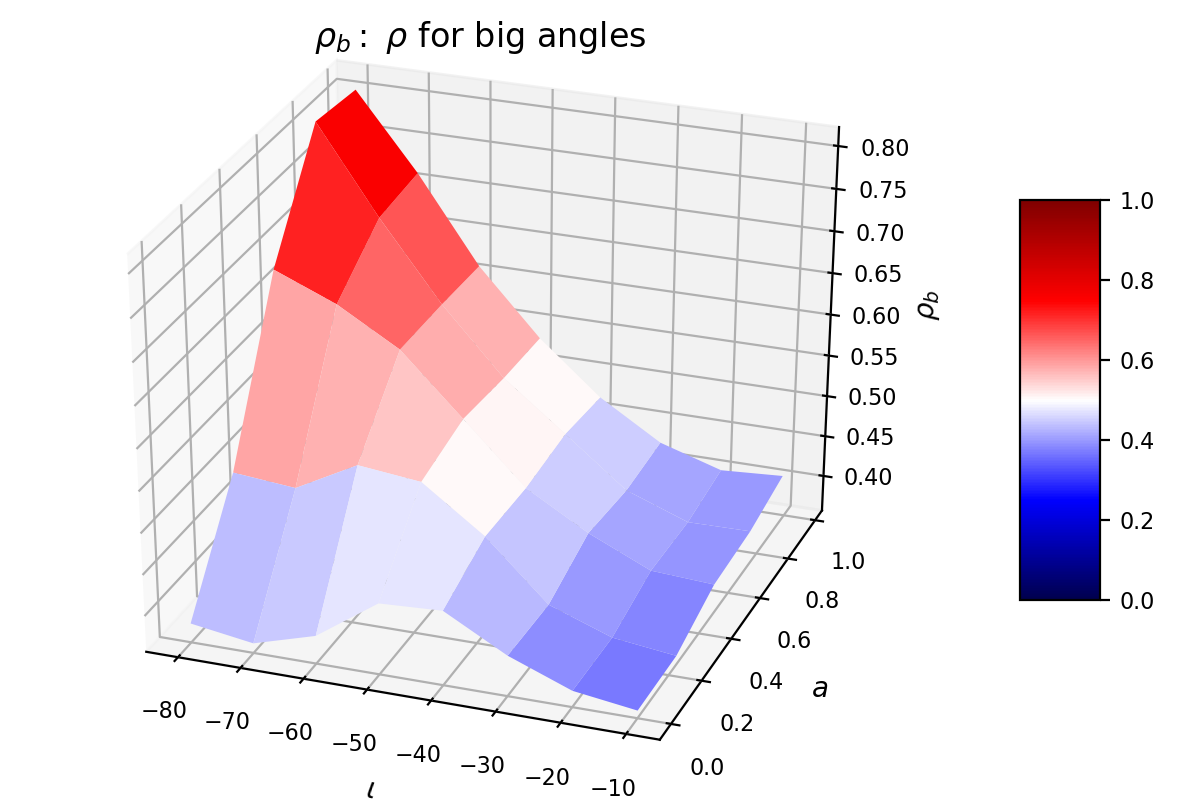}
\caption{Comparison of the above graphs, with big angles, against the
	EHT image, through the calculation of the correlation, as explained in the text.
}
\label{fig:rho-b}
\end{figure*}

It can be seen that the position of the lump, in our images, with the highest intensity
does not coincide with none of the three most intense zones, seen in the
elliptical EHT image.
For this reason in the next subsection we study images where the projection
of the angular momentum of the black hole, coincides with the expected
direction for the angular momentum of the galaxy.


\subsection{Graphs with angular momentum aligned with the angular momentum of the galaxy}\label{subsec:ang-galax}

The rather high values 
of the correlation coefficient
shown in Fig. \ref{fig:rho-b}
is not convincing evidence for us
that the synthetic images
are giving a good representation
of the EHT image. In fact, one
can clearly notice that the morphology of them are very different,
and this is a quality that
is not meassured by the
correlation coefficient.

For this reason we have also considered 
the case in which the plane of the disk
has small angles with respect to
the plane of the galaxy.
We use the following values of iota:
$\iota = -0.0873$, 
$\iota = -0.0300$, 
$\iota = -0.0030$, 
$\iota = -0.0003$, 
$\iota = -0.000001$, 
$\iota =  0.000001$, 
$\iota =  0.0003$, 
$\iota =  0.0030$, 
$\iota =  0.0300$, 
$\iota =  0.0873$;
corresponding to degrees of
$-5^o$, 
$-1.72^o$, 
$-0.172^o$, 
$-0.0172^o$, 
$-0.00005729^o$, 
$ 0.00005729^o$, 
$ 0.0172^o$, 
$ 0.172^o$, 
$ 1.72^o$, 
$ 5^o$, 
of the angle of the black hole angular momentum with the plane of the image.

\begin{figure*}
\centering
\includegraphics[clip,width=0.195\textwidth]{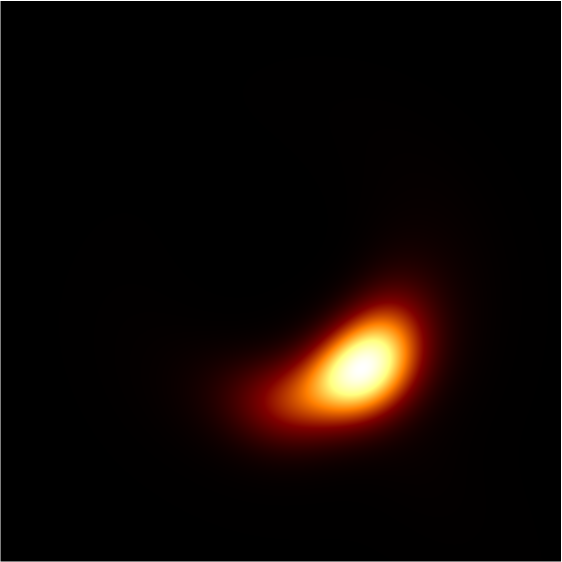}
\includegraphics[clip,width=0.195\textwidth]{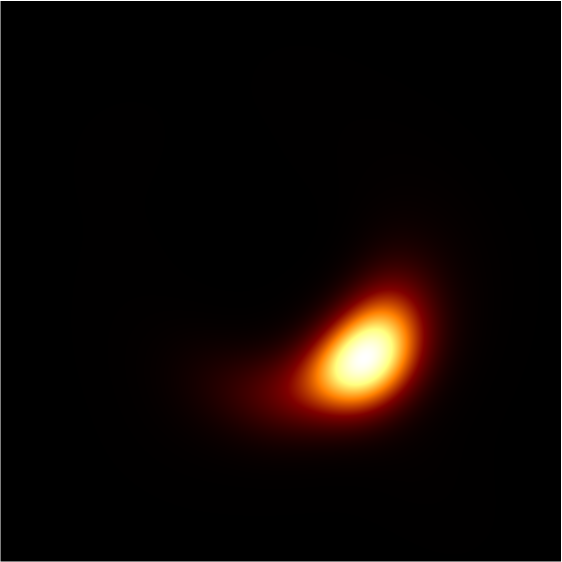}
\includegraphics[clip,width=0.195\textwidth]{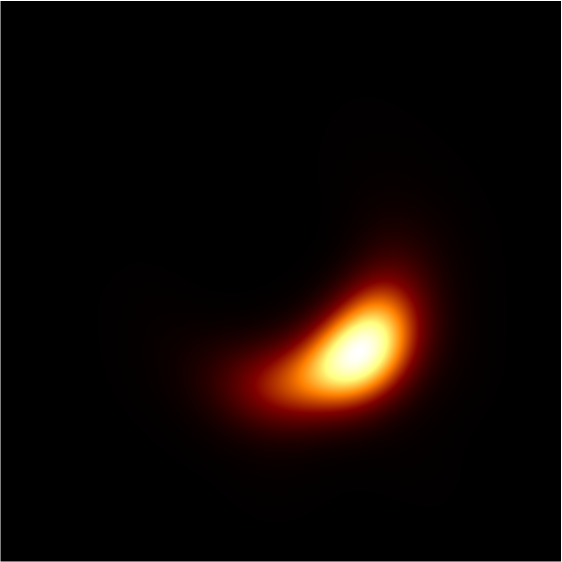}
\includegraphics[clip,width=0.195\textwidth]{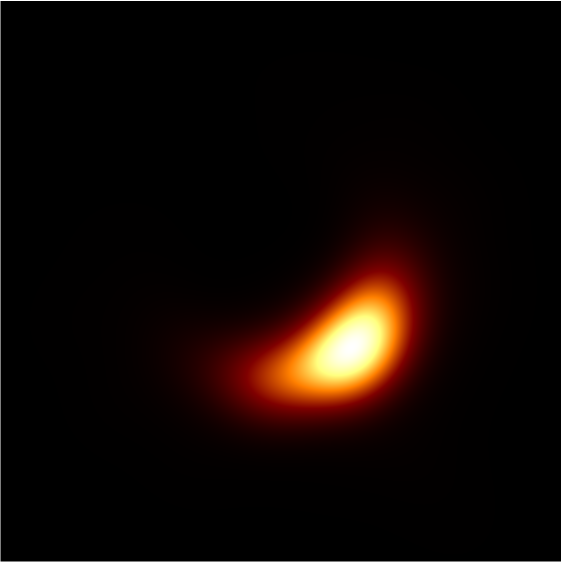}
\includegraphics[clip,width=0.195\textwidth]{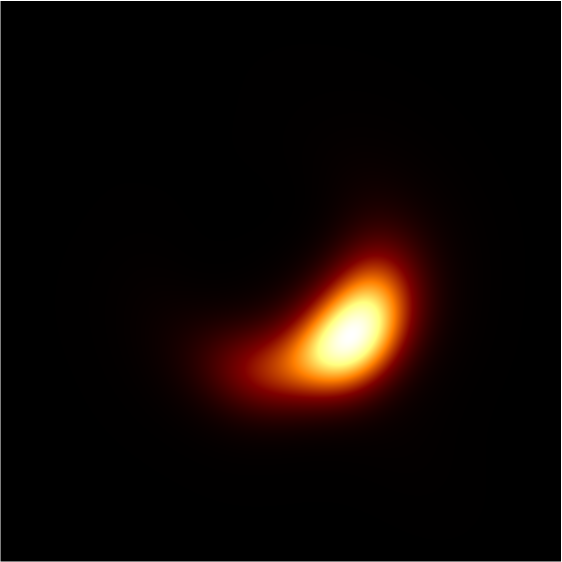}
\caption{$\iota=-0.0873=-5^\circ$, and from left to right with Kerr parameter: 
		$a=0.00$,  $a=0.25$, $a=0.50$, $a=0.75$ and $a=0.98$.
}
\label{fig:io-0873}
\end{figure*}

\begin{figure*}
\centering
\includegraphics[clip,width=0.195\textwidth]{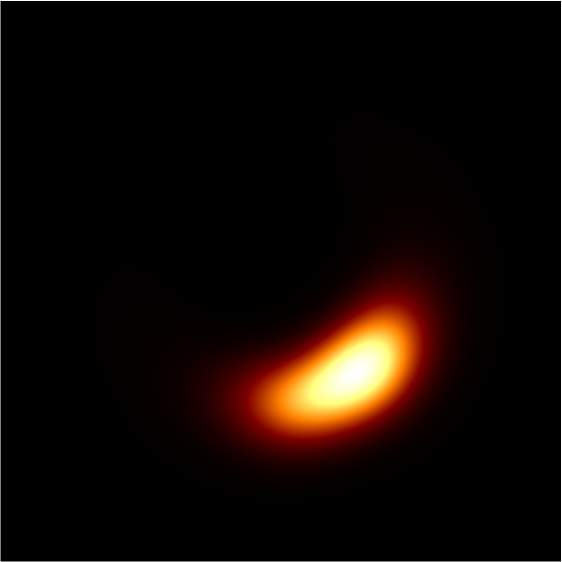}
\includegraphics[clip,width=0.195\textwidth]{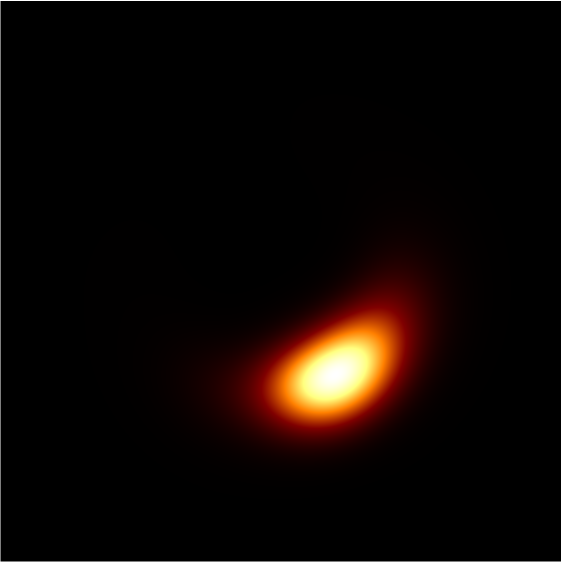}
\includegraphics[clip,width=0.195\textwidth]{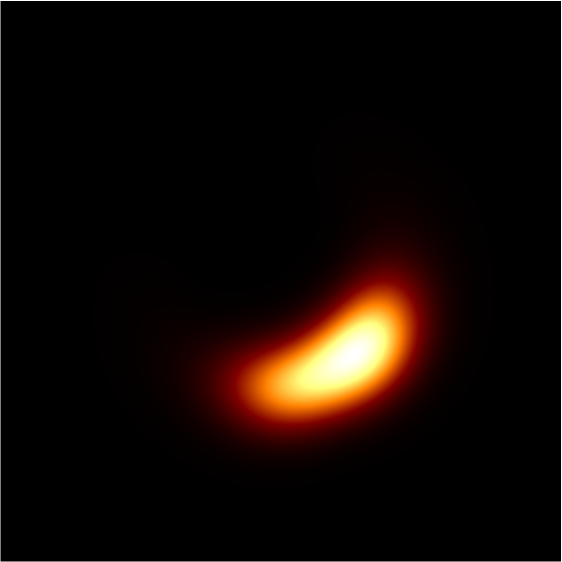}
\includegraphics[clip,width=0.195\textwidth]{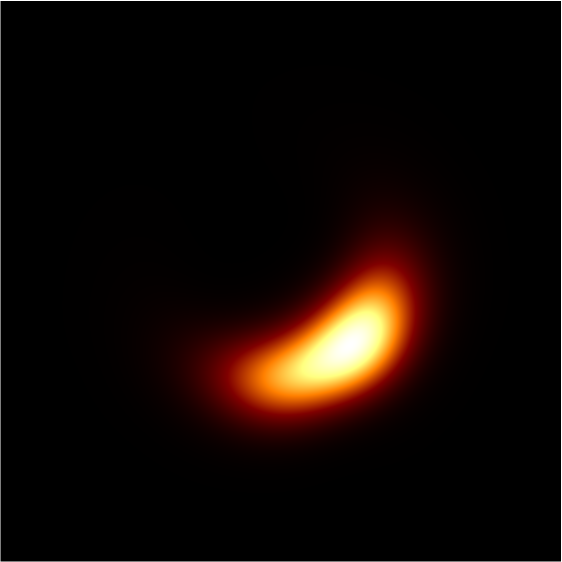}
\includegraphics[clip,width=0.195\textwidth]{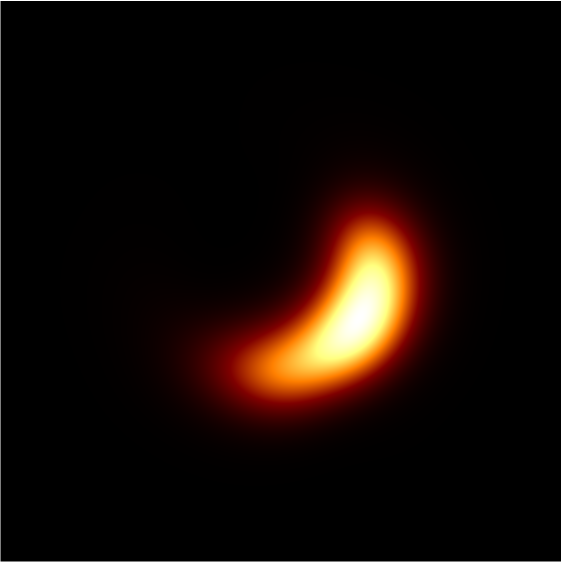}
\caption{$\iota=-0.0300=-1.72^\circ$, and from left to right with Kerr parameter: 
		$a=0.00$,  $a=0.25$, $a=0.50$, $a=0.75$ and $a=0.98$.
}
\label{fig:io-0300}
\end{figure*}

\begin{figure*}
\centering
\includegraphics[clip,width=0.195\textwidth]{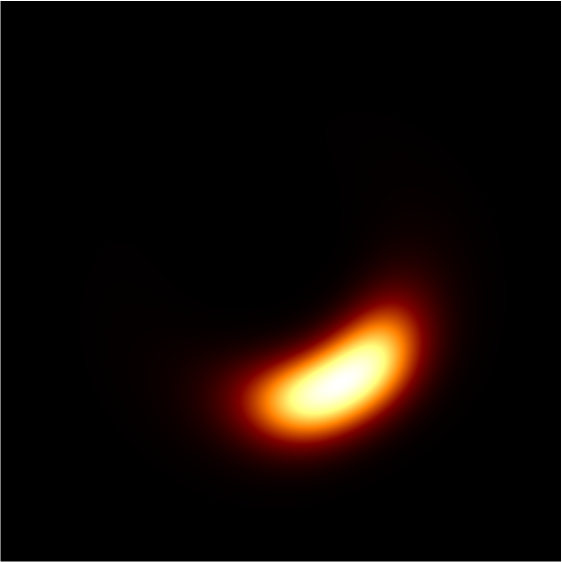}
\includegraphics[clip,width=0.195\textwidth]{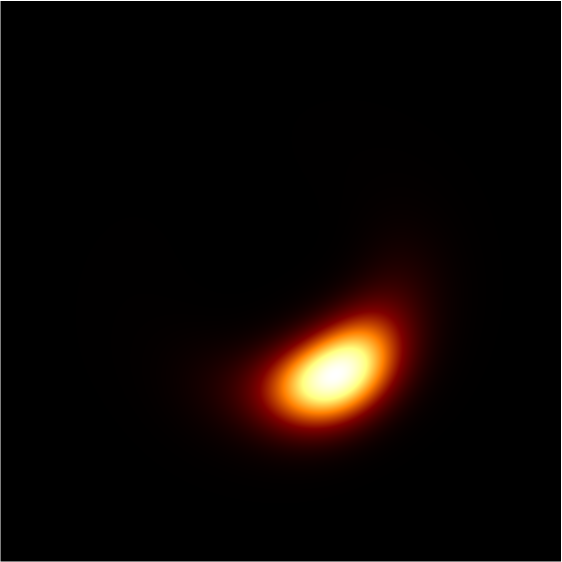}
\includegraphics[clip,width=0.195\textwidth]{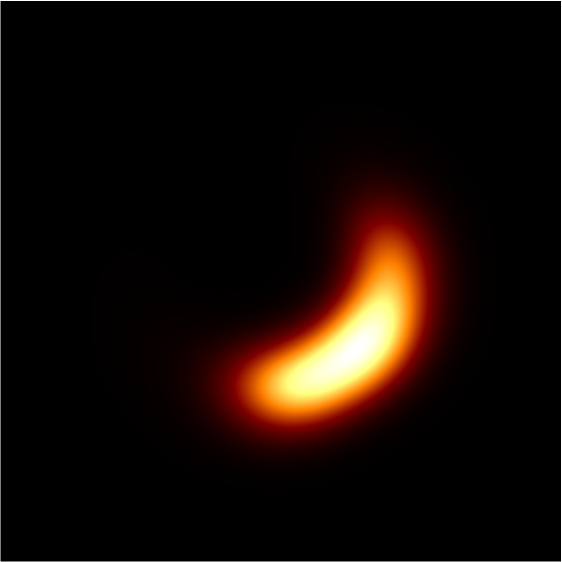}
\includegraphics[clip,width=0.195\textwidth]{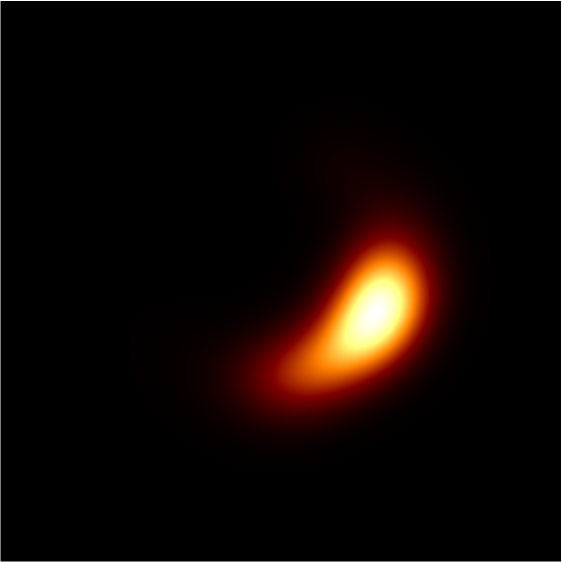}
\includegraphics[clip,width=0.195\textwidth]{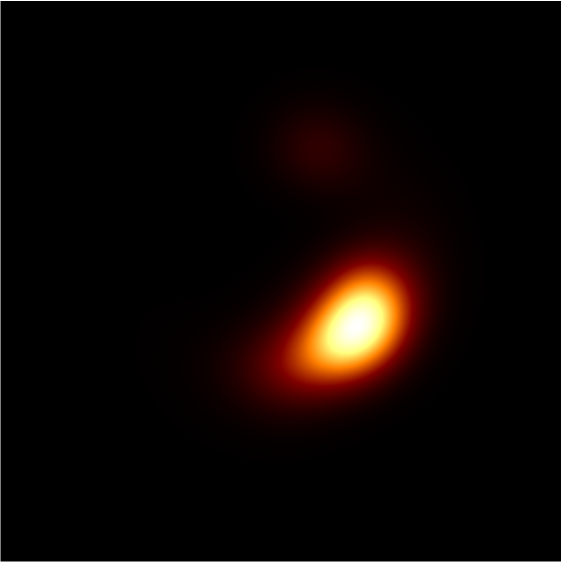}
\caption{$\iota=-0.0030=-0.172^\circ$, and from left to right with Kerr parameter: 
	$a=0.00$,  $a=0.25$, $a=0.50$, $a=0.75$ and $a=0.98$.
}
\label{fig:io-0030}
\end{figure*}

\begin{figure*}
\centering
\includegraphics[clip,width=0.195\textwidth]{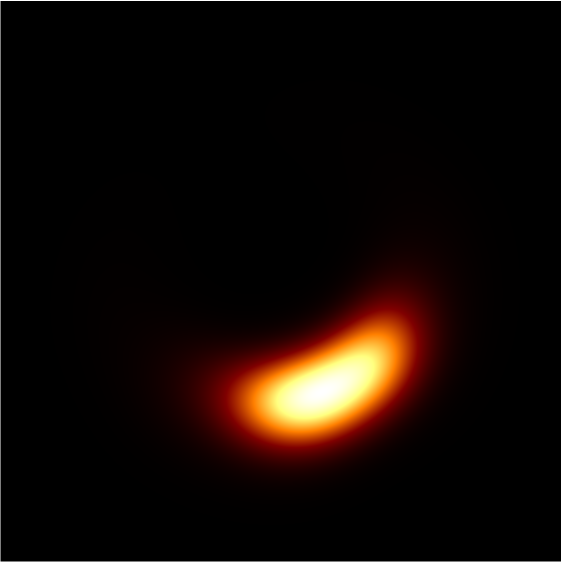}
\includegraphics[clip,width=0.195\textwidth]{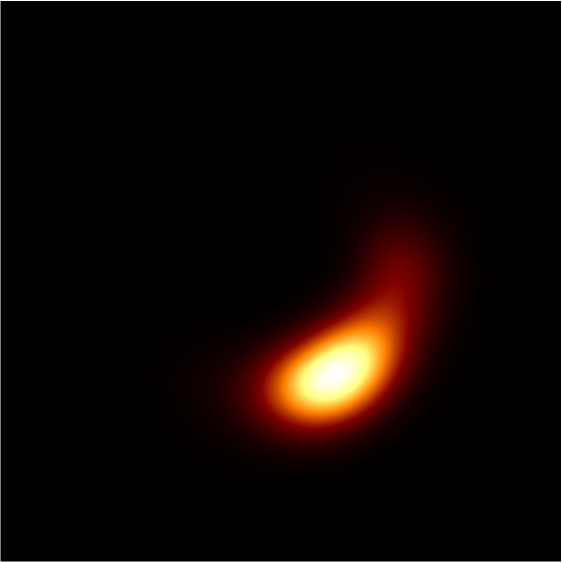}
\includegraphics[clip,width=0.195\textwidth]{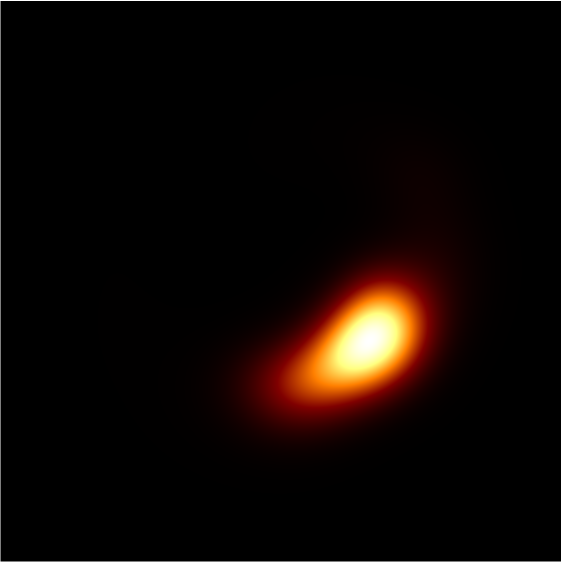}
\includegraphics[clip,width=0.195\textwidth]{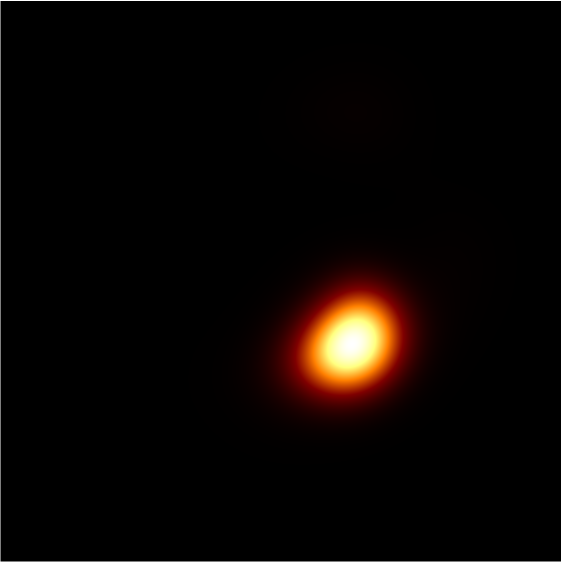}
\includegraphics[clip,width=0.195\textwidth]{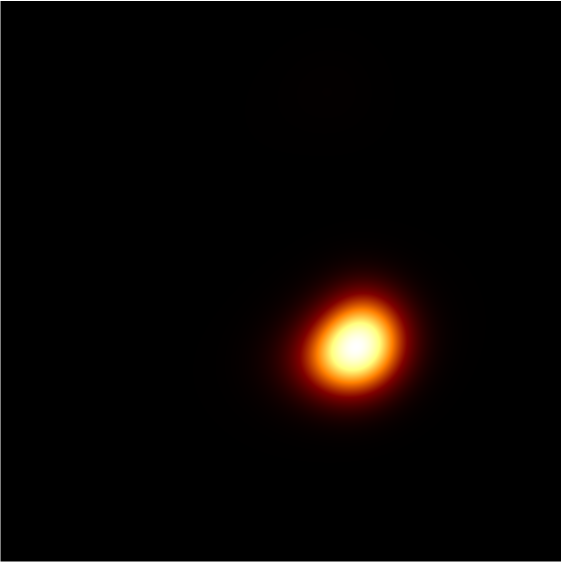}
\caption{$\iota=-0.0003=-0.0172^\circ$, and from left to right with Kerr parameter: 
		$a=0.00$,  $a=0.25$, $a=0.50$, $a=0.75$ and $a=0.98$.
}
\label{fig:io-0003}
\end{figure*}

\begin{figure*}
\centering
\includegraphics[clip,width=0.195\textwidth]{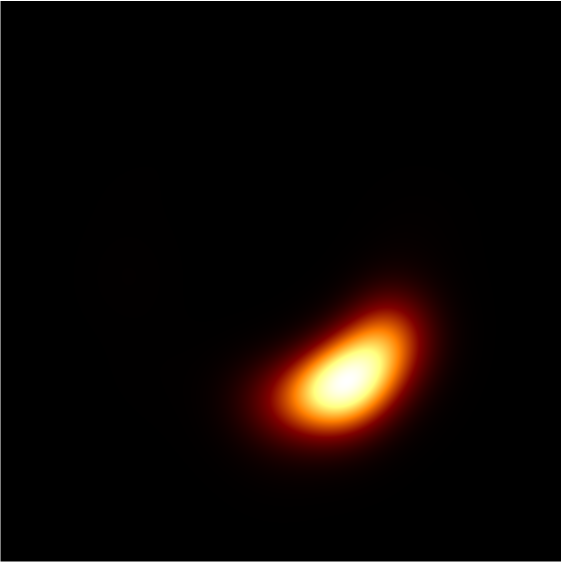}
\includegraphics[clip,width=0.195\textwidth]{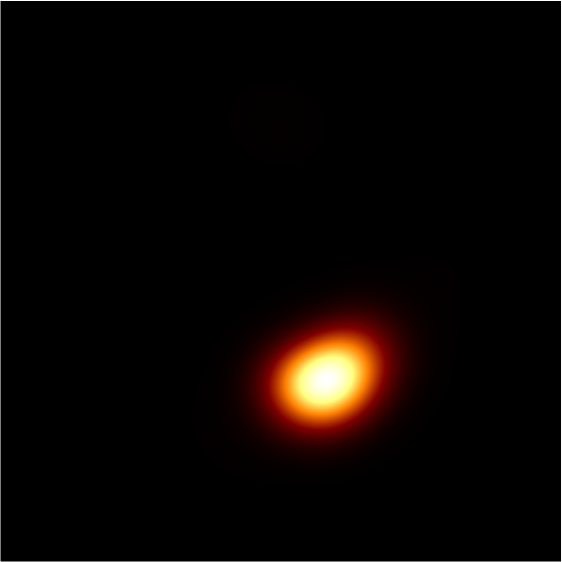}
\includegraphics[clip,width=0.195\textwidth]{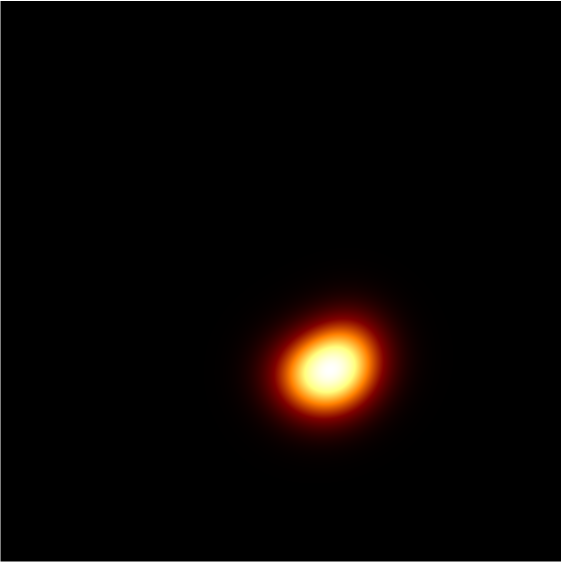}
\includegraphics[clip,width=0.195\textwidth]{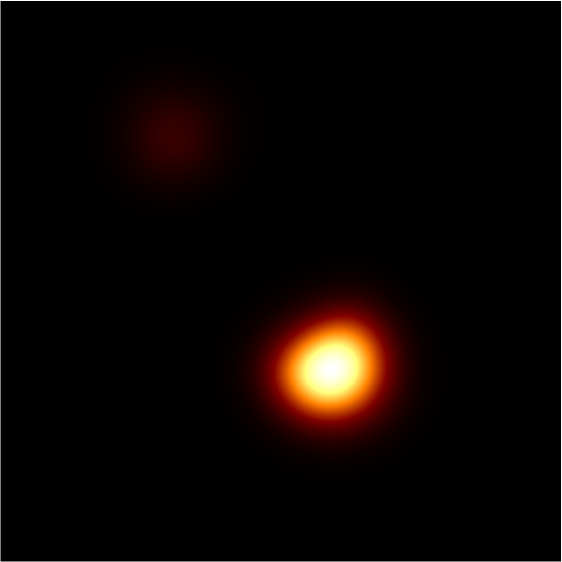}
\includegraphics[clip,width=0.195\textwidth]{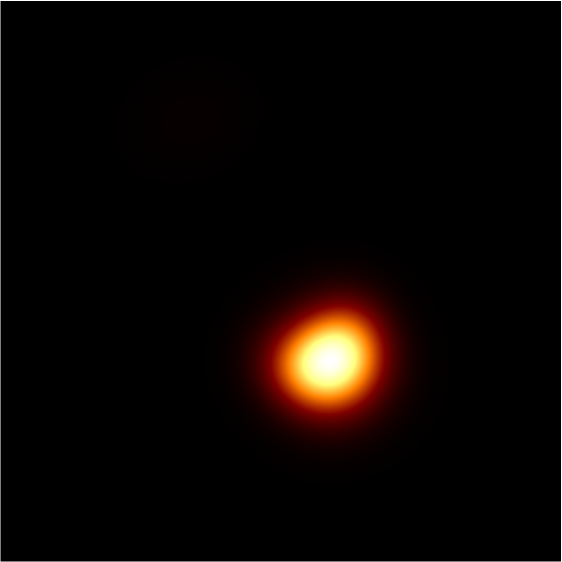}
\caption{$\iota=-10^{-6}=-0.00005729^\circ$, and from left to right with Kerr parameter: 
		$a=0.00$,  $a=0.25$, $a=0.50$, $a=0.75$ and $a=0.98$.
}
\label{fig:io-000001}
\end{figure*}

\begin{figure*}
\centering
\includegraphics[clip,width=0.195\textwidth]{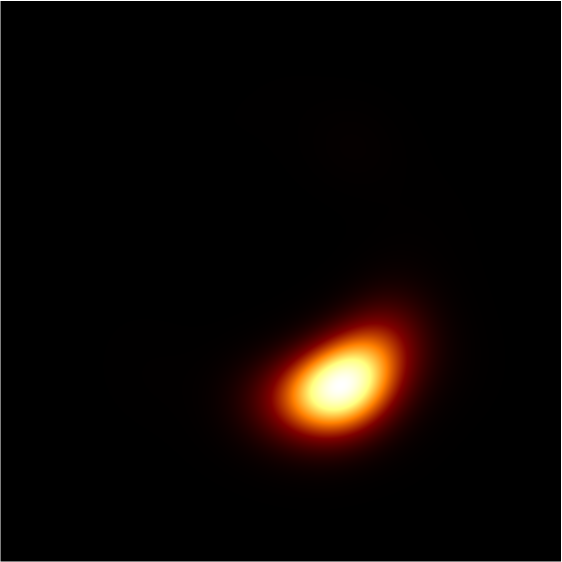}
\includegraphics[clip,width=0.195\textwidth]{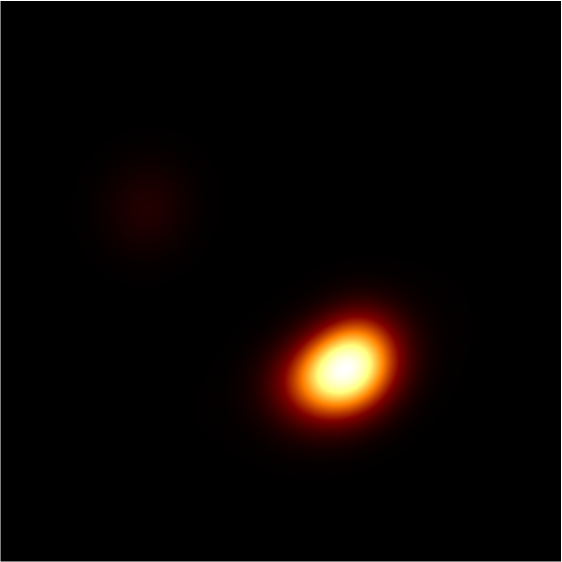}
\includegraphics[clip,width=0.195\textwidth]{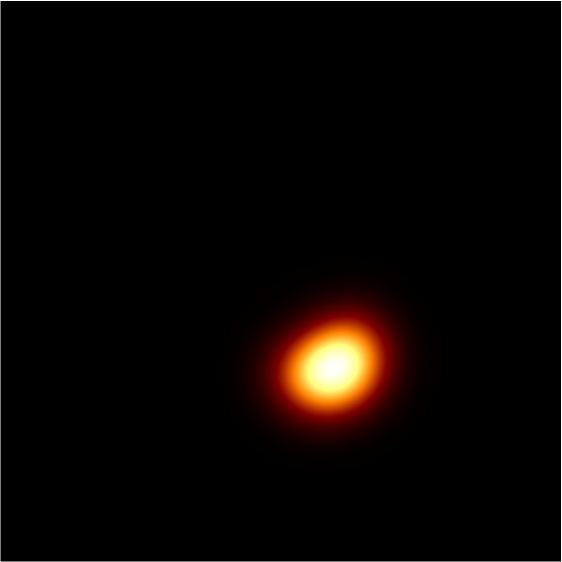}
\includegraphics[clip,width=0.195\textwidth]{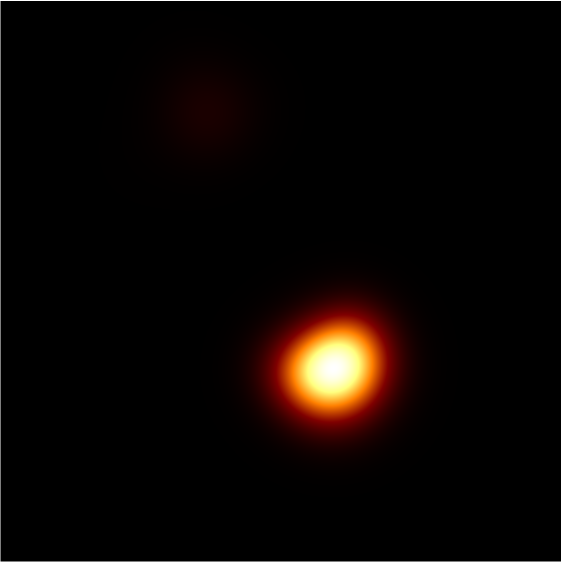}
\includegraphics[clip,width=0.195\textwidth]{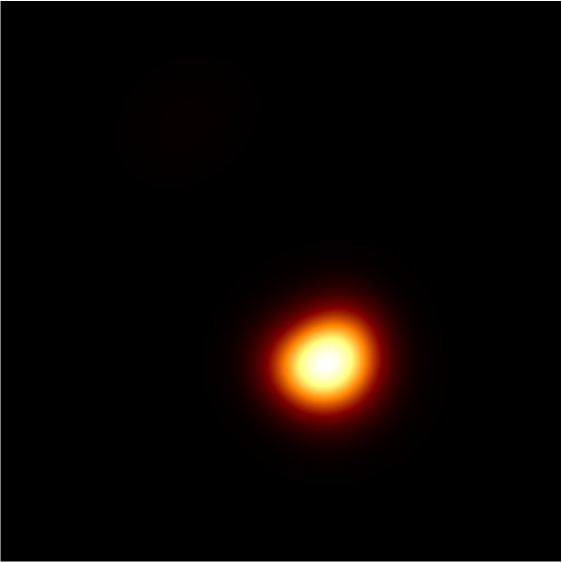}
\caption{$\iota=10^{-6}=0.00005729^\circ$, and from left to right with Kerr parameter: 
		$a=0.00$,  $a=0.25$, $a=0.50$, $a=0.75$ and $a=0.98$.
}
\label{fig:io000001}
\end{figure*}

\begin{figure*}
\centering
\includegraphics[clip,width=0.195\textwidth]{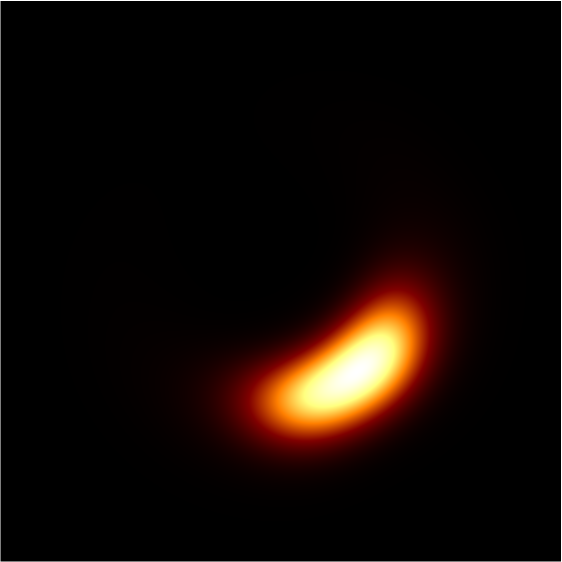}
\includegraphics[clip,width=0.195\textwidth]{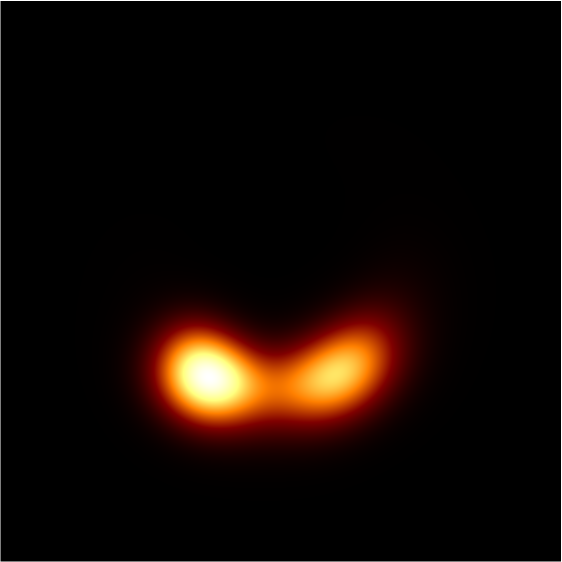}
\includegraphics[clip,width=0.195\textwidth]{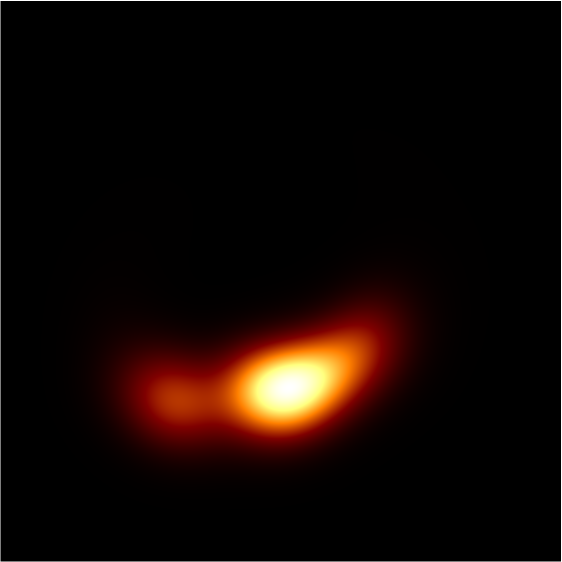}
\includegraphics[clip,width=0.195\textwidth]{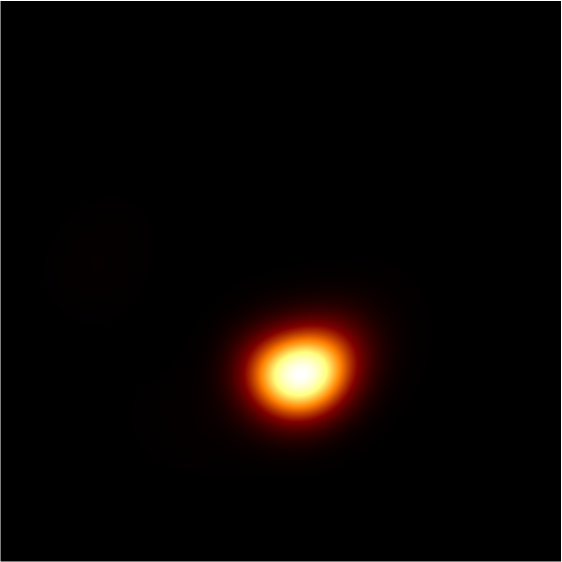}
\includegraphics[clip,width=0.195\textwidth]{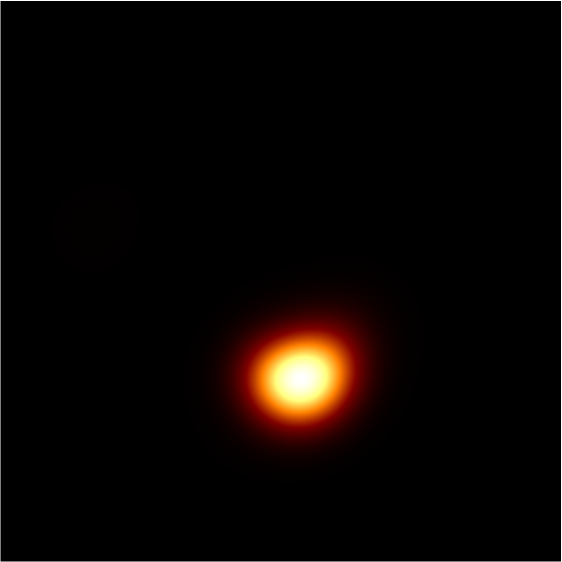}
\caption{$\iota=0.0003=0.0172^\circ$, and from left to right with Kerr parameter: 
		$a=0.00$,  $a=0.25$, $a=0.50$, $a=0.75$ and $a=0.98$.
}
\label{fig:io0003}
\end{figure*}

\begin{figure*}
\centering
\includegraphics[clip,width=0.195\textwidth]{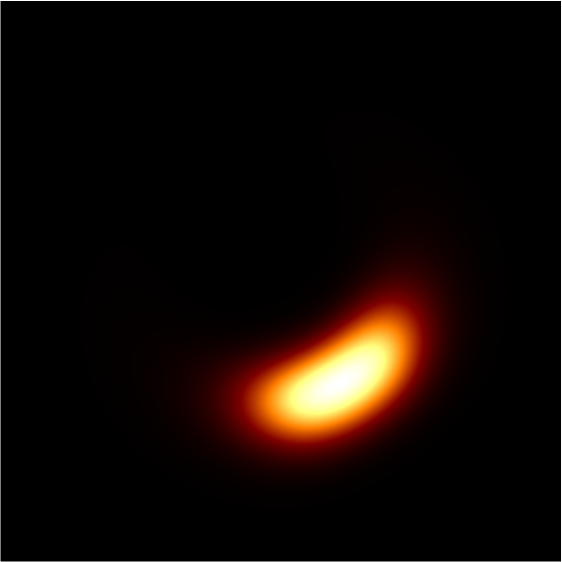}
\includegraphics[clip,width=0.195\textwidth]{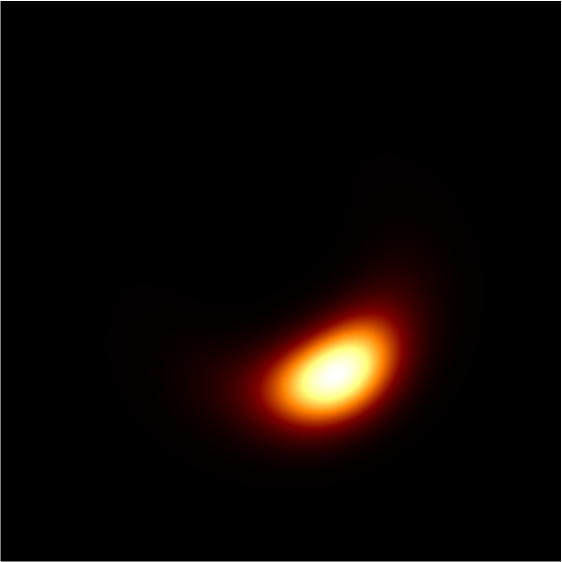}
\includegraphics[clip,width=0.195\textwidth]{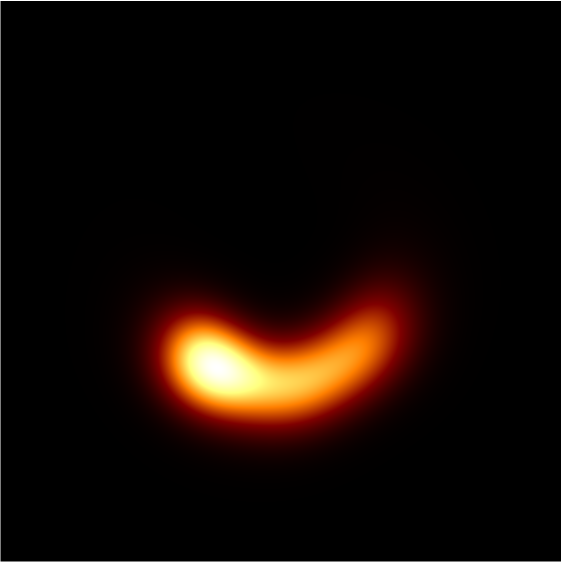}
\includegraphics[clip,width=0.195\textwidth]{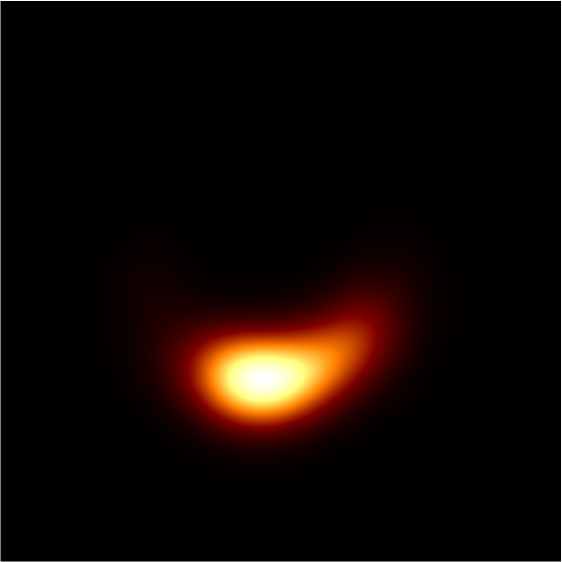}
\includegraphics[clip,width=0.195\textwidth]{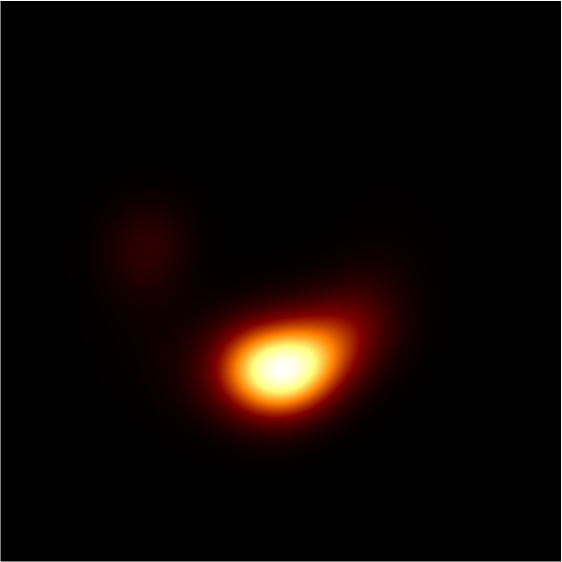}
\caption{$\iota=0.0030=0.172^\circ$, and from left to right with Kerr parameter: 
		$a=0.00$,  $a=0.25$, $a=0.50$, $a=0.75$ and $a=0.98$.
}
\label{fig:io0030}
\end{figure*}

\begin{figure*}
\centering
\includegraphics[clip,width=0.195\textwidth]{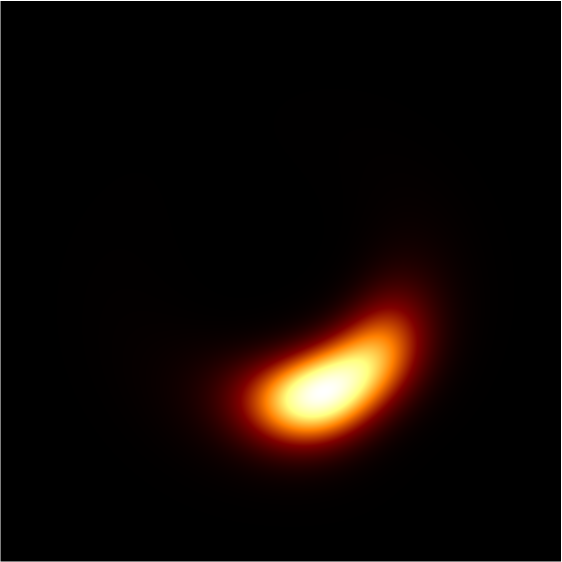}
\includegraphics[clip,width=0.195\textwidth]{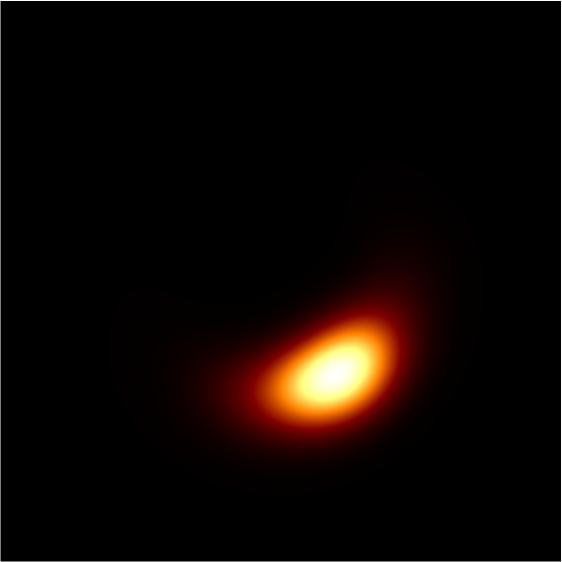}
\includegraphics[clip,width=0.195\textwidth]{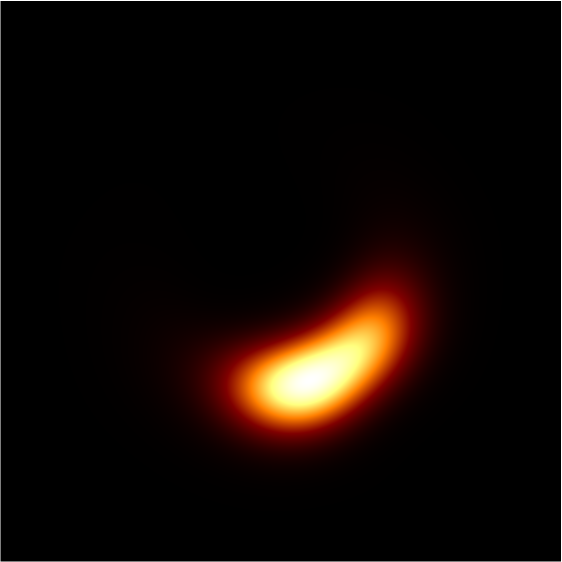}
\includegraphics[clip,width=0.195\textwidth]{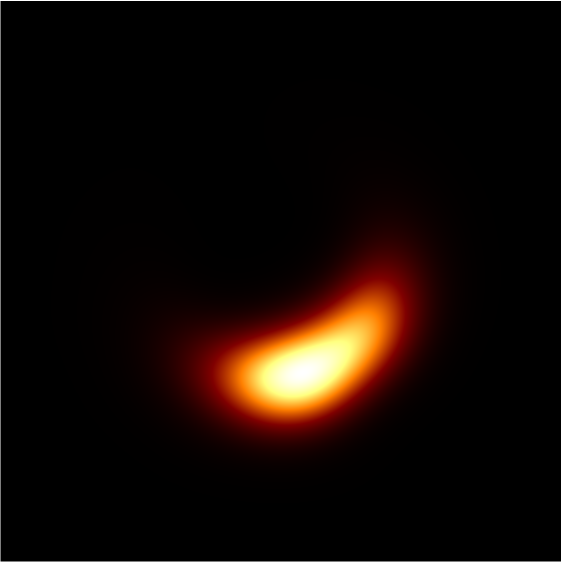}
\includegraphics[clip,width=0.195\textwidth]{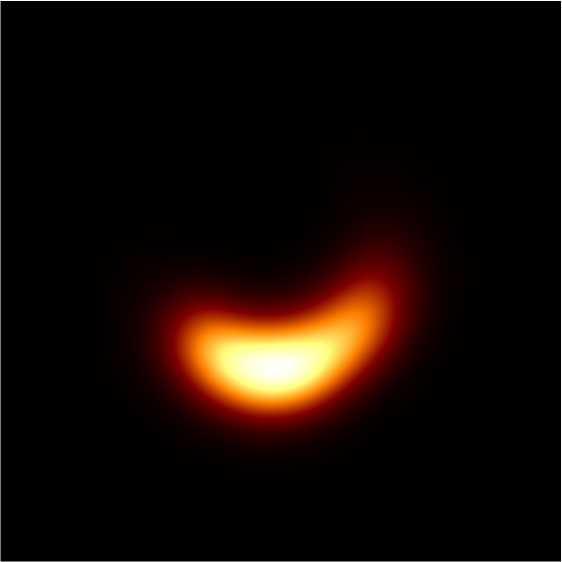}
\caption{$\iota=0.0300=1.72^\circ$, and from left to right with Kerr parameter: 
	$a=0.00$,  $a=0.25$, $a=0.50$, $a=0.75$ and $a=0.98$.
}
\label{fig:io0300}
\end{figure*}

\begin{figure*}
\centering
\includegraphics[clip,width=0.195\textwidth]{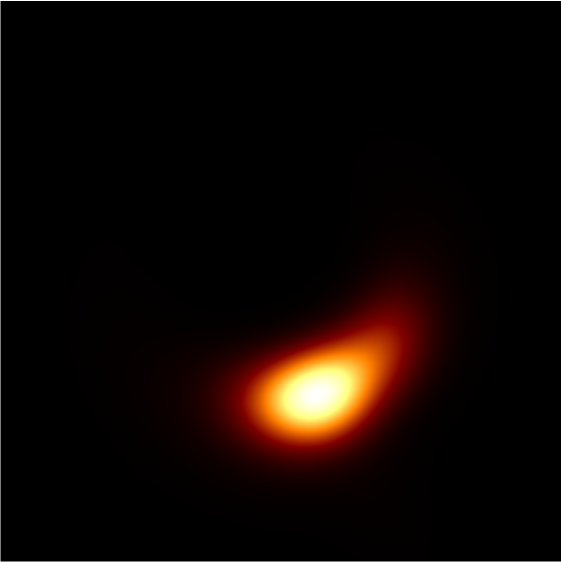}
\includegraphics[clip,width=0.195\textwidth]{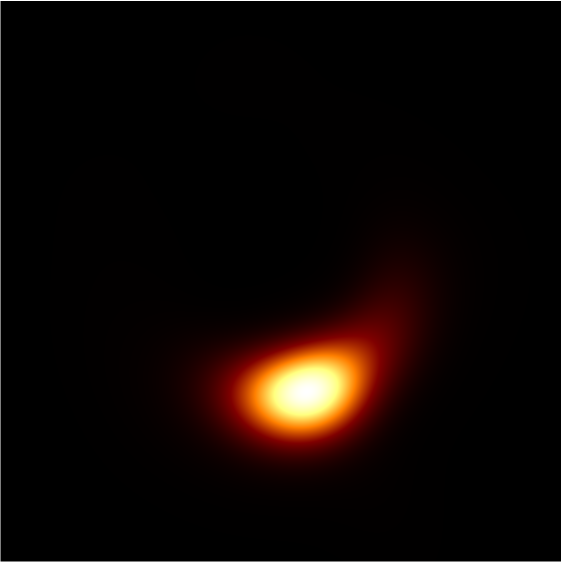}
\includegraphics[clip,width=0.195\textwidth]{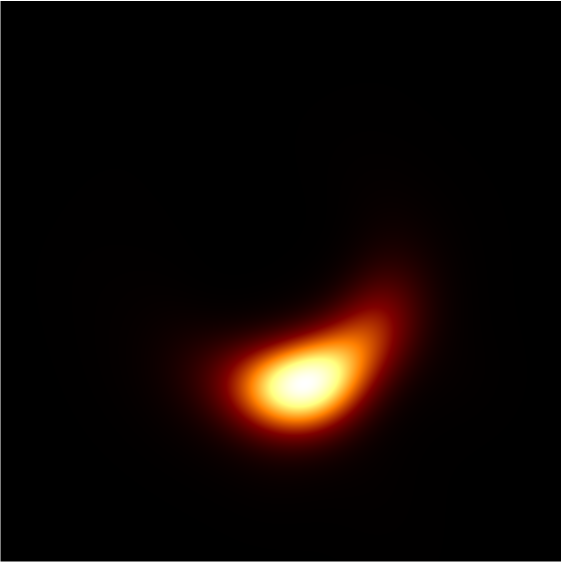}
\includegraphics[clip,width=0.195\textwidth]{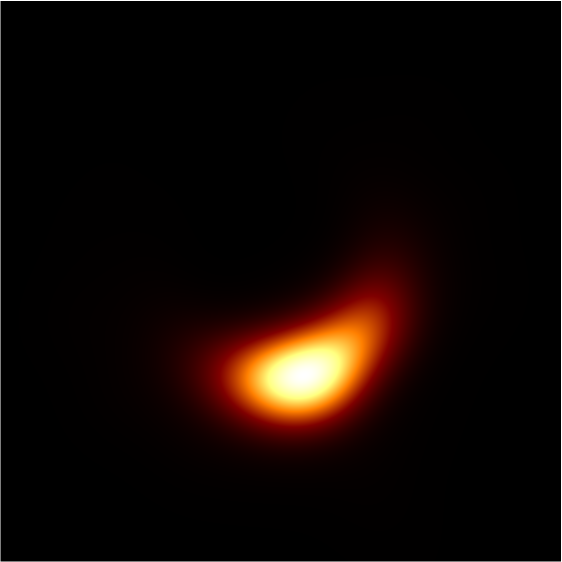}
\includegraphics[clip,width=0.195\textwidth]{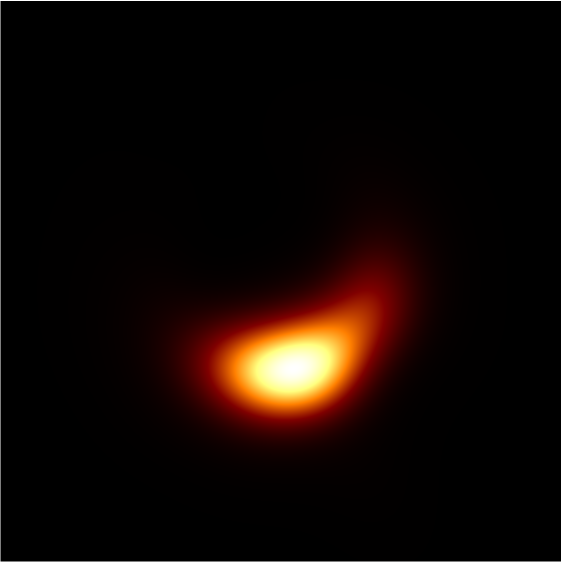}
\caption{$\iota=0.0873=5^\circ$, and from left to right with Kerr parameter: 
		$a=0.00$,  $a=0.25$, $a=0.50$, $a=0.75$ and $a=0.98$.
}
\label{fig:io0873}
\end{figure*}

From figure \ref{fig:io-0873} to \ref{fig:io0873} we show the graphs corresponding to images
calculated with an angular momentum of the black hole, whose projection
is approximately
that of the angular momentum of the galaxy.
From left to right, in each figure we show the images corresponding to 
$a=0.00$,  $a=0.25$, $a=0.50$, $a=0.75$ and $a=0.98$ respectively.

It should be notice that due to the small inclination of the 
disc, simulated images require a very good control of numerical errors since magnification effects becomes divergent near the 
emission disc region.


\begin{figure*}
\centering
\includegraphics[clip,width=0.7\textwidth]{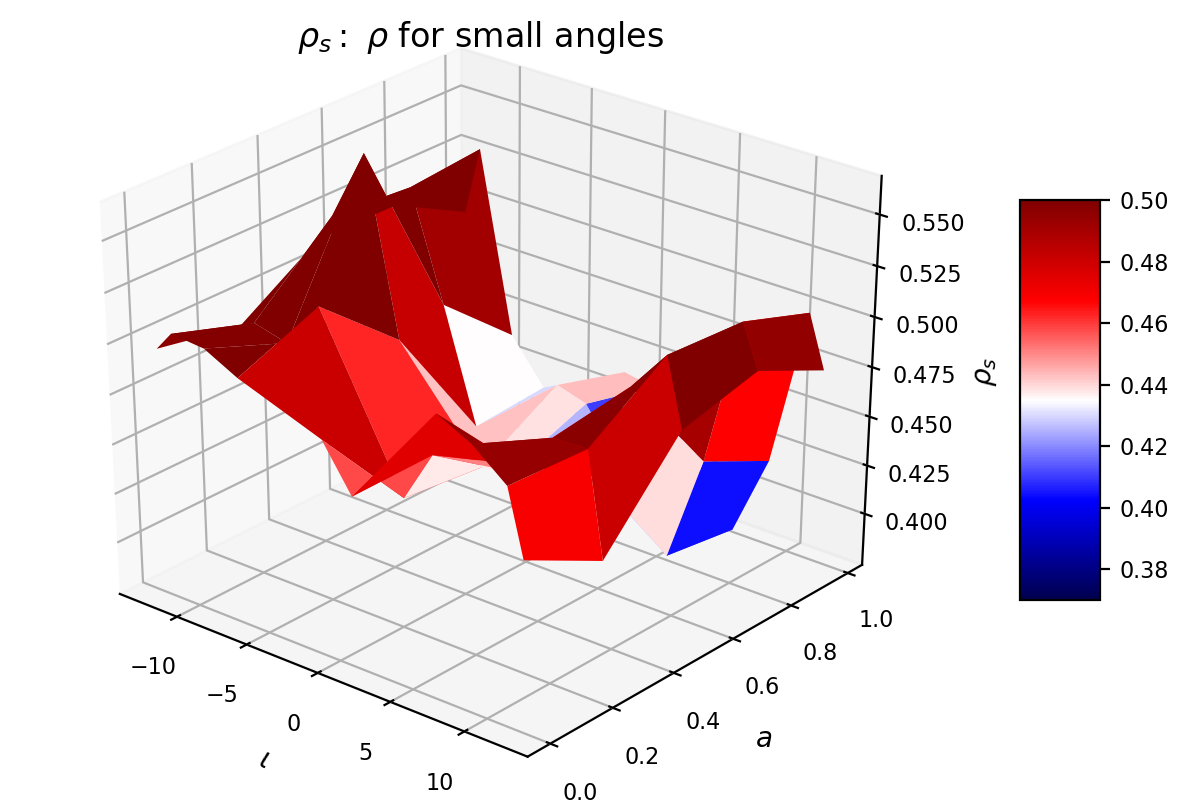}
\caption{Comparison of the above graphs, with small angles, against the
	EHT image, through the calculation of the correlation, as explained in the text.
	The $\iota$ axis has the logarithmic scale defined by $\ln(\iota e 10^6)$, where $e$ is Euler number;
	so that the small angles are magnified. 
}
\label{fig:rho-s}
\end{figure*}
For convenience we also present the values of the correlations that produces the graph of Fig. \ref{fig:rho-s}
in table \ref{tab:rho-s}.
\begin{table}
\begin{center}
\begin{tabular}{c||c|c|c|c|c|}  
& 0.00 & 0.25 & 0.50 & 0.75 & 0.98 \\
\hline
$-5^o$  & 0.4969 & 0.4891 & 0.5144 & 0.5140 & 0.4946 \\
$-1.72^o$ & 0.5069 & 0.4937 & 0.5289 & 0.5256 & 0.5290 \\
$-0.172^o$ & 0.5054 & 0.4894 & 0.5646 & 0.4715 & 0.4388 \\
$-0.0172^o$ & 0.4968 & 0.5131 & 0.4779 & 0.4143 & 0.4169 \\
$-0.00005729^o$ & 0.4816 & 0.4324 & 0.4279 & 0.4505 & 0.4395 \\
$ 0.00005729^o$ & 0.4589 & 0.4593 & 0.4312 & 0.4460 & 0.4368 \\
$ 0.0172^o$ & 0.5142 & 0.4698 & 0.4438 & 0.3825 & 0.3771 \\
$ 0.172^o$ & 0.5057 & 0.4889 & 0.4944 & 0.4375 & 0.4194 \\
$ 1.72^o$& 0.4919 & 0.4896 & 0.5156 & 0.5130 & 0.5003 \\
$ 5^o$ & 0.4590 & 0.4376 & 0.4818 & 0.4915 & 0.4743 \\
\hline
\end{tabular}
\end{center}
\caption{Table of values of the correlation coefficients for small angles,
for different values of the angular momentum parameter
in columns,
as shown in Fig. \ref{fig:rho-s}.}\label{tab:rho-s}
\end{table}
\begin{figure*}
\centering
\includegraphics[clip,width=0.195\textwidth]{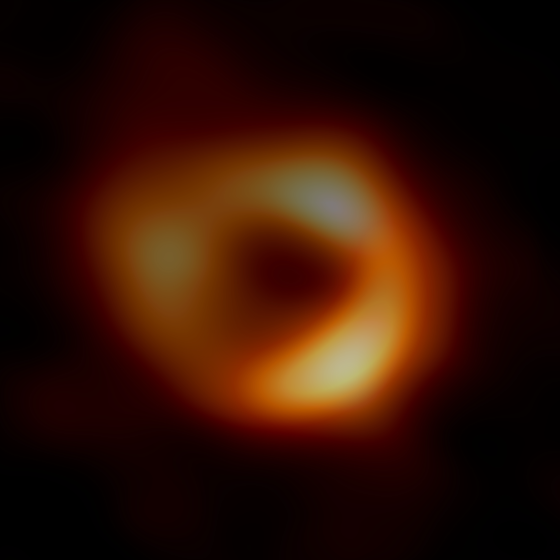}
\includegraphics[clip,width=0.195\textwidth]{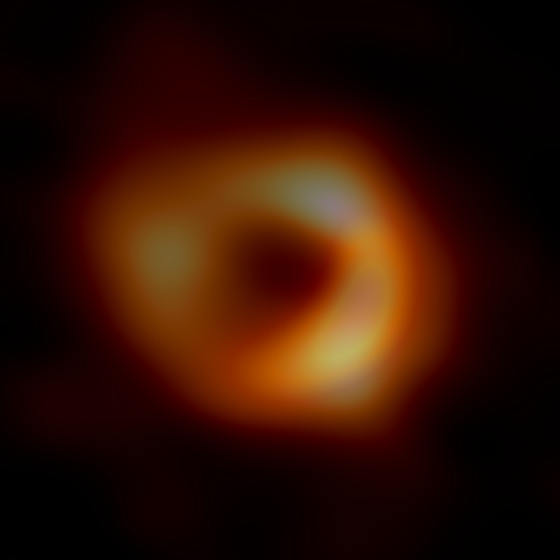}
\includegraphics[clip,width=0.195\textwidth]{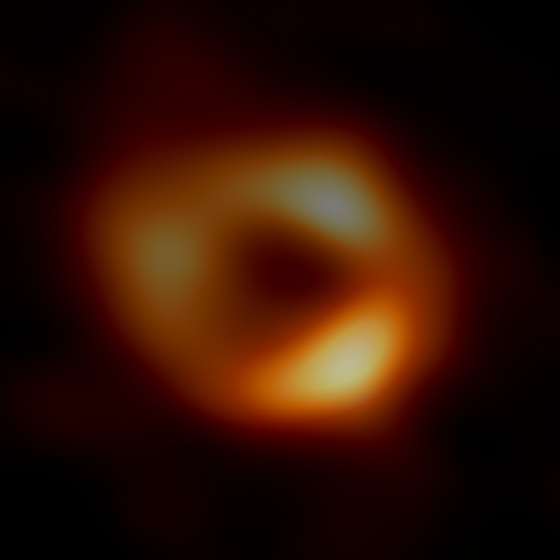}
\caption{Superposition of best candidates with the original image.
	These are from left to right, the cases: 
	$a=0.50$ with $\iota = -0.003$,
	$a=0.98$ with $\iota = -0.030$
	and
	$a=0.50$ with $\iota = -0.030$.
	It can be seen that the first case shows the best match of the
	local maximum at the lower right of the image.
}
\label{fig:superposition}
\end{figure*}

\subsection{End point of null geodesics in Kerr-Schild coordinates}

We have presented two sets of simulated images using as a physical model
that of a thin disk with different equatorial inclinations with respect
to the line of sight, and also for various values of the angular momentum
parameter $a$.
For reasons that are being discussed, we take as our best model,
the one characterized by the values $a=0.5$ and $\iota=-0.003$.
But the question we would like to tackle now, is, where do all
these ray tracing calculation reach, in the vicinity of the black hole?
The question is relevant to understand the nature of the image
one is generating.
The first thing to clarify is the choice of a coordinate system
to use, in order to make graphs that have some geometrical meaning
that is not corrupted by a bad behavior of the coordinate system;
as is known to happen with the Boyer-Lindquist coordinates in the vicinity
of the event horizon.
Our choice is the Kerr-Schild\citep{Kerr-Schild-1965} coordinate system, that is naturally
defined when one makes the corresponding decomposition of the Kerr metric,
in the so called Kerr-Schild form.
In this way we have at our disposal a set of coordinates 
$(\mathsf{t},x,y,z)$ that we use for these graphs.

\begin{figure*}
\centering
\includegraphics[clip,width=0.49\textwidth]{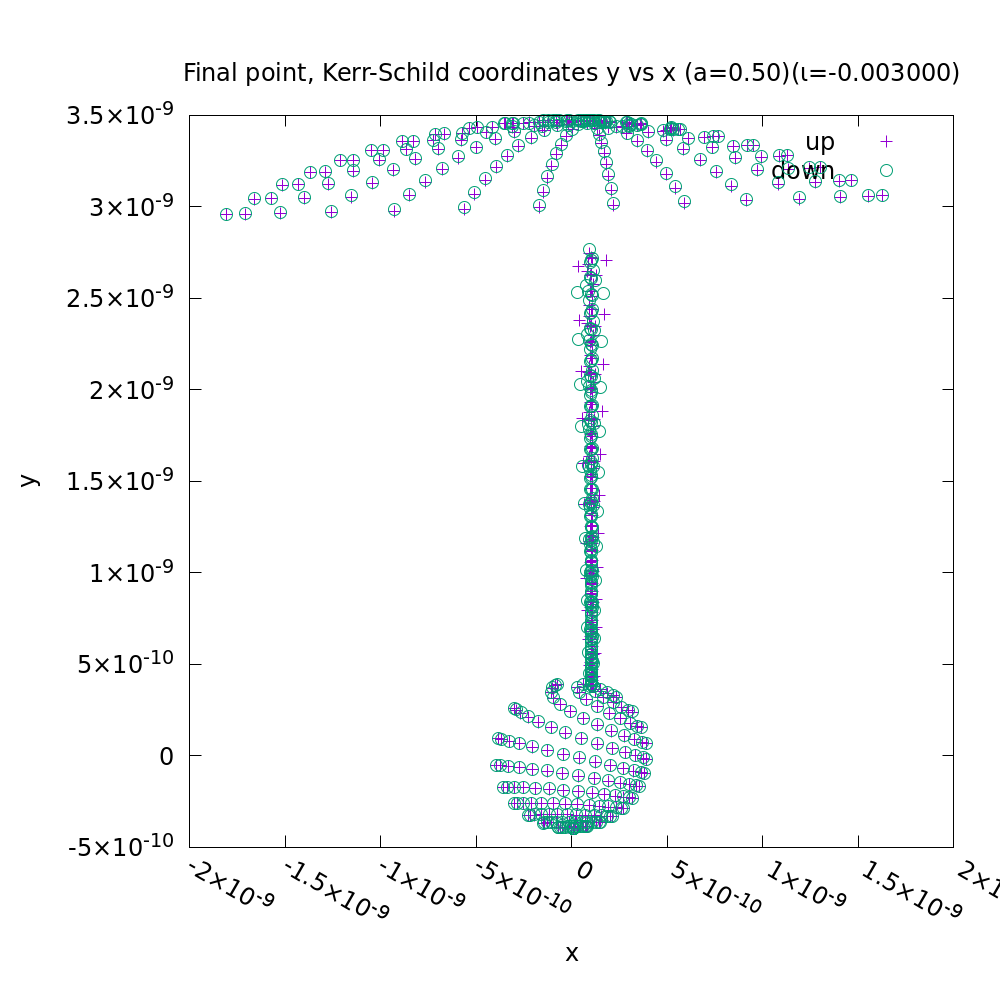}
\includegraphics[clip,width=0.49\textwidth]{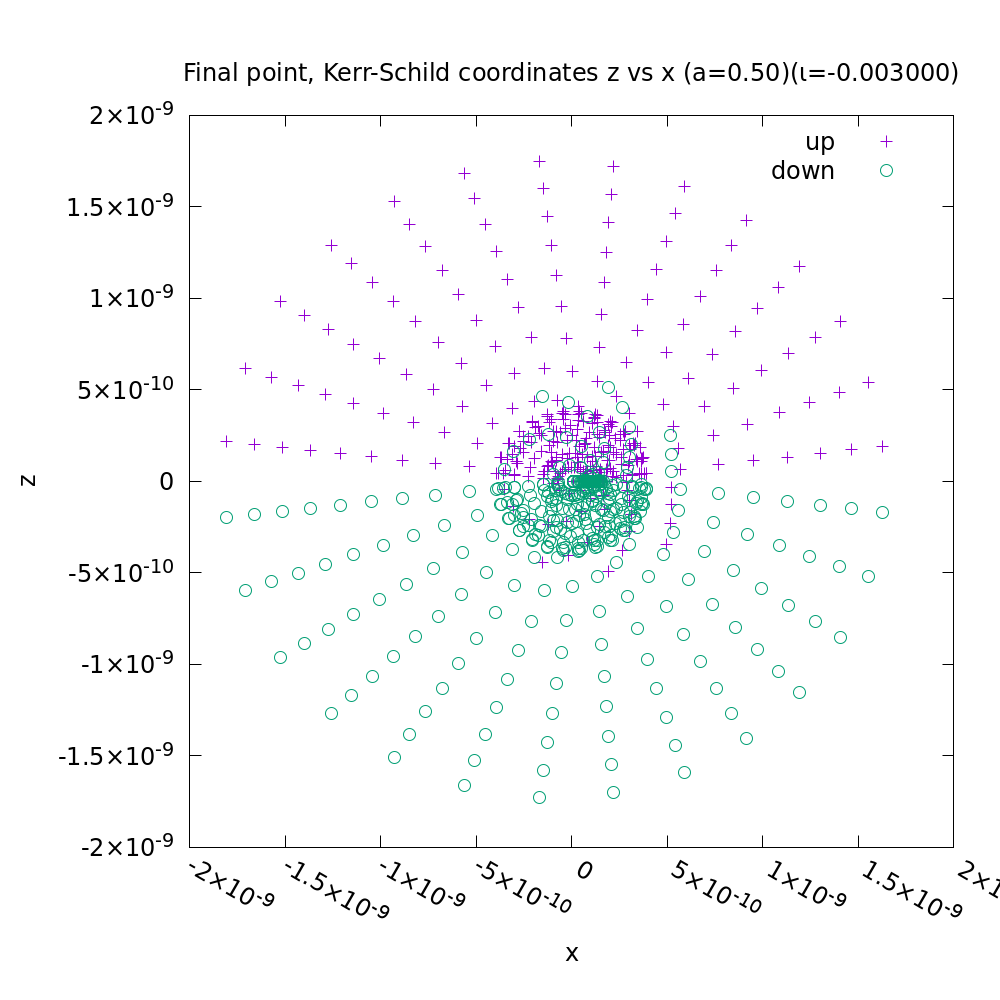}
\includegraphics[clip,width=0.49\textwidth]{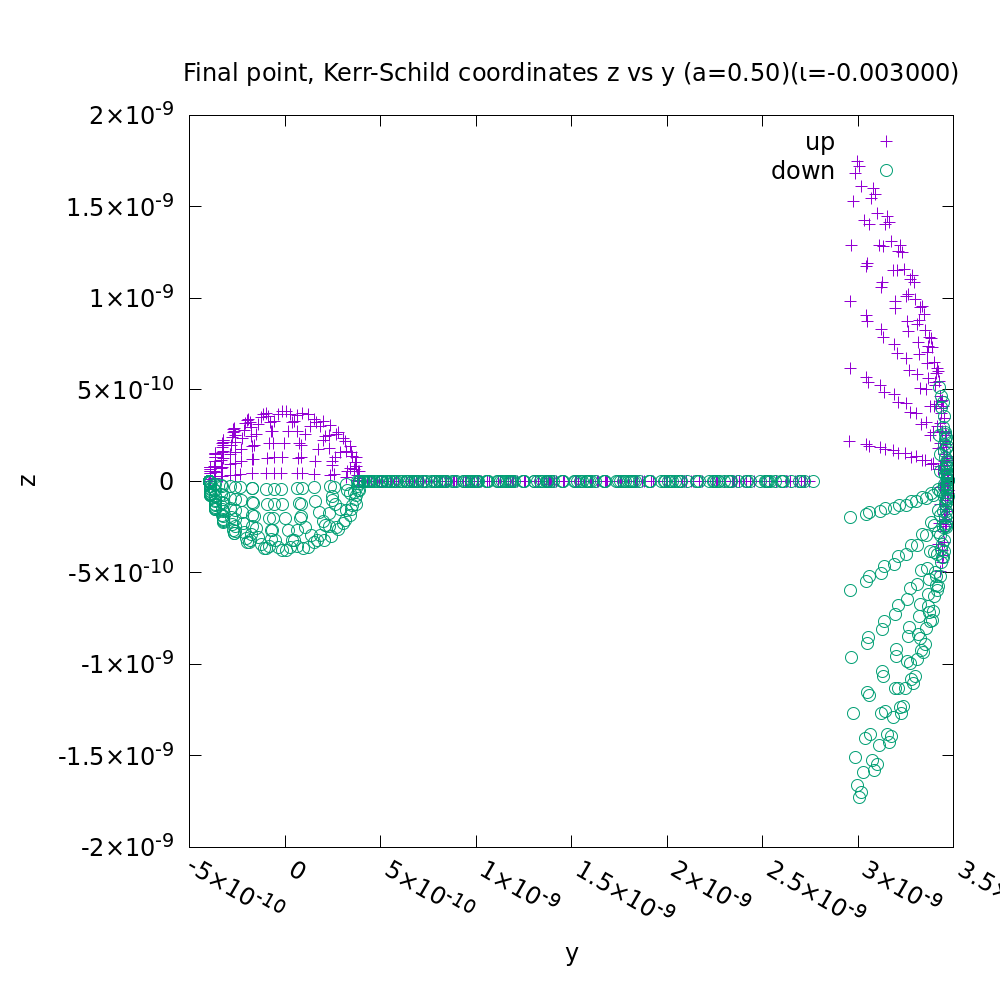}
\caption{The Kerr-Schild coordinates are adapted to the geometry of the black hole,
	so that the equatorial plane is $z=0$, and the angular momentum of the black hole
	points in the direction of the positive $z$'s.
	The top-left graph, using $(x,y)$ coordinates, shows the end points of the ray tracing to
	the past, for all values of $z$; where positive $y$ means the zone away from the observer,
	that is, in the opposite side of the black hole.
The top-right graph, using $(x,z)$ coordinates, shows the end points of the null geodesics to
the past, for all values of $y$, 
the bottom  graph, using $(y,z)$ coordinates, shows the end points of the ray tracing to
the past, in this plane for all values of $x$.
}
\label{fig:kerrshild-a05-io1e-3}
\end{figure*}
In the graphs of Fig. \ref{fig:kerrshild-a05-io1e-3}
we present the end points of the integration
in three different planes.
In the top-left graph, it is impressive to see that those trajectories that reach
the disk, are almost aligned along the line in the equatorial plane, that
is opposite to the line of sight. That is, one has to have in mind the the observer
in these coordinates is located somewhere far away with big values of the negative
$y$ coordinate. In other words, with these geometrical settings, the observer only
sees a very small portion of the disk which is just opposite to the line of sight;
since for this small angle, the front part is almost invisible; in fact, non
of our geodesics reach any other region of the disk.
This is in spite of the fact that we have made the calculation for 
1008 geodesics in the plane of the expected image.
The upper part of this graph, shows an umbrella type shape,
which just shows our choice for the end condition, for those
geodesic that have not reached the disk.
In the bottom part of the same graph, one can see an sphere like object,
which is just the condition that the geodesics have reached 
the event horizon.
For this rather low angular momentum, one can only see a peculiar
spread of the end points; where in the whole graph,
crosses indicates geodesics that start above the equatorial plane
and circles denote geodesics that start below the equatorial plane.
The fact that the spread in the disk is so small, also indicates
the degree of precision of our calculations.
In particular we have been using quadruple precision
with $2.4 \times 10^{-22}$ tolerance for the 
Runge-Kutta integrator\citep{rksuite-90}.

In the top-right graph, showing the $(x,z)$ values,
one can see the umbrella type shape, from another angle.
A big number of end points, at the opposite side of the black hole,
are difficult to see in this graph, since they are confused
with the end points of null geodesics that reach
the event horizon, and form the sphere like shape
object, more or less in the center of this graph.

Finally in the bottom graph of Fig. \ref{fig:kerrshild-a05-io1e-3},
one can see the same distribution from a different angle;
in this case the $(x,z)$ coordinates.
Since the equatorial plane coincides with $z=0$, one can see
that the condition for the null geodesics to reach the disk
is satisfied with very high precision;
which are all those points outside the event horizon
with $z=0$.
The umbrella type shape is now shown to the right of this graph.

\subsection{The possible bar structure on the disk}

In our previous article on the construction of synthetic images of M87\citep{Boero:2021},
we recurred to a bar structure on the disk in order to build
images that resemble in a better way the corresponding EHT image of 
the supermassive black hole.
For this reason, we have also study for this case of Sgr~A${}^*$,
the consequences of introducing this type of structure.
The results did not help in providing better images, that
where closer to the aspects found in the EHT image of Sgr~A${}^*$.
And the reason for this, can be understood in terms of the graphs
of Fig. \ref{fig:kerrshild-a05-io1e-3}, discussed previously.
That is, since for this geometric configuration, the only
observed part of the disk lies on a narrow sector, on the disk,
at the opposite region of the line of sight; the introduction
of a bar structure with higher temperature, can only be seen
if it is placed on this narrow region. And therefore
its effect is to slightly change the original image,
without bar structure.
Consequently we have not included those graphs in this
article, since they do not contribute to the search
for the explanation of the other structure shown
in the EHT image of Fig. \ref{fig:eht-image}.

\subsection{Numerical comparison of the images}

The main part of the construction process of the images,
is the analysis of the general aspects of EHT image.
Guided by the suggested geometry of Fig. \ref{fig:eht-image},
we have made a couple of configuration for our disk model,
in order to try to construct images that resemble
that of the EHT team.
But it is convenient to also have at hand some numerical measure
for the comparison of our images with that of EHT.
For this reason, as we did in our previous work, we have
also considered the correlation
between two images. This is a very limited type of measure,
that does not include information on the structure of 
each image; but in any case it is a very natural type of measure.

The comparison of our first configuration, with big angles
between the equatorial plane of the disk and the plane of the galaxy
are shown graphically in Fig. \ref{fig:rho-b}.
While, the comparison of our second configuration, with small angles
between the equatorial plane of the disk and the plane of the galaxy
are shown graphically in Fig. \ref{fig:rho-s}.
The corresponding table for this last graph is presented in table \ref{tab:rho-s};
where one can see that our chosen configuration of $a=0.5$ and $\iota=-0.003$,
has the highest value.

\section{Photon regions and silhouette of the black hole}
\label{sec:silueta}

\begin{figure*}
\centering
\includegraphics[clip,width=0.6\textwidth]{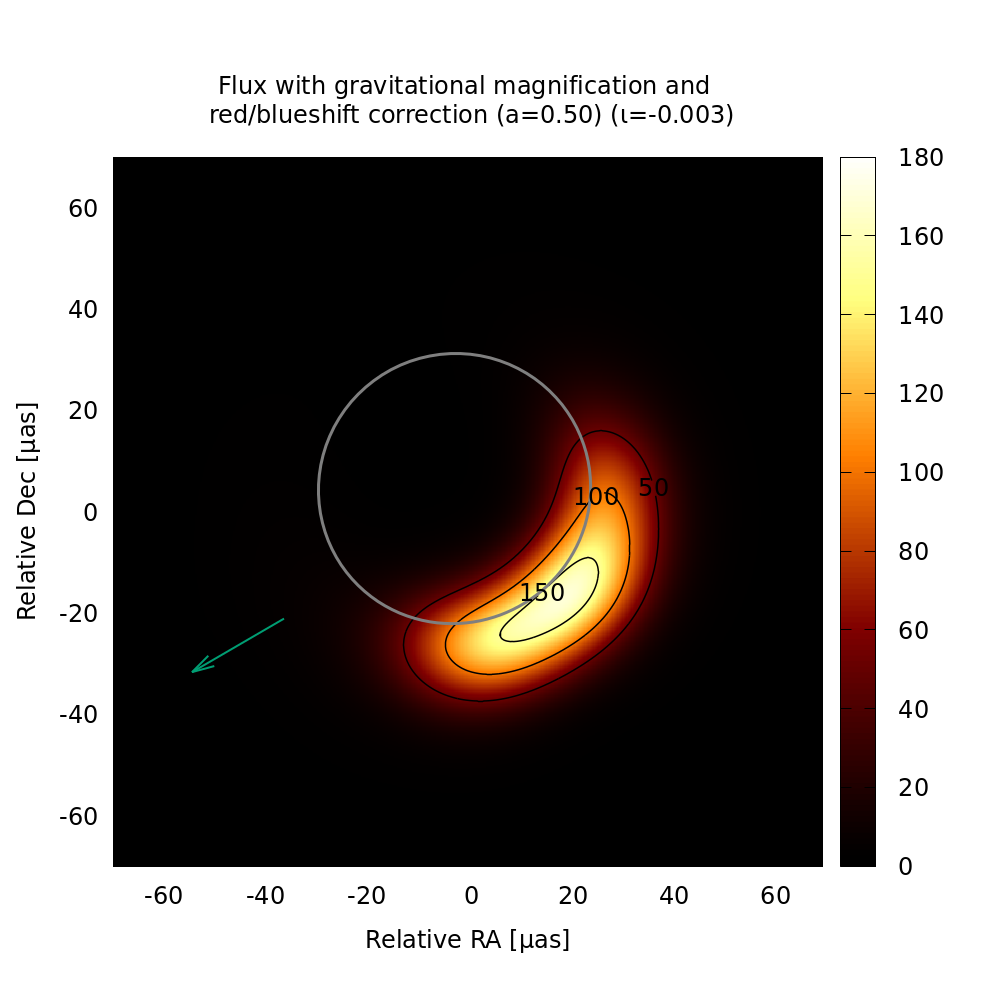}
\caption{Image of the flux with all magnifications for Sgr~A${}^*$
    in the case of $a=0.50$ and $\iota=-0.003$; where the grey line shows the silhouette
    for the case of a back ilumination of the supermassive black hole.
}
\label{fig:image-with-silhouette}
\end{figure*}

In this section we introduce the necessary algebra
needed for the calculation of the silhouette of
a black hole with angular momentum that is tilted
with respect to the line of sight,
and that we show in Fig. \ref{fig:image-with-silhouette}.

In a spherically symmetric spacetime it is natural to study the null geodesics
that have constant radial coordinate; and so define the corresponding photon region.
Surprisingly, in Kerr spacetime, the analog question also has meaning.
It should be noticed that since the spacetime is not spherically symmetric,
it is not trivial to ask whether there exists a natural radial coordinate;
but Kerr's radial coordinate\cite{Kerr63} $r$ seems to be the answer.
In fact, there exist null geodesics with $r=$constant, and they have been called
also the generators of the `photon regions'.
However only recently there have been claims\cite{Cederbaum:2019vwv}
that this set contains the only trajectories of photons that do not go cross
the horizon or escape to infinity.
In relation to all this, it is probably worthwhile to point out that
the radial and angular coordinates $(r,\theta)$ used
in the Boyer-Lindquist\cite{Boyer67} form of the metric,
coincide with the original $(r,\theta)$ presented in Kerr's article.

Although the photon regions have been studied in several works, 
we are only concerned here with the silhouette that an observer
would see of a black hole with angular momentum, if seen with
a lighted background.

Then, we are concerned with the null geodesics that satisfy
$r=r_p$, and so,
$\dot{r}=0$ and $\ddot{r}=0$; or equivalently,
$\mathcal{R}=0$ and $\frac{d\mathcal{R}}{dr}=0$,
which can be expressed as:
\begin{equation}
0 = \Big( E\left(r_p^2 + a^2 \right) - a L \Big)^2 - K \Delta(r_p) 
,
\label{eq:R+geod+func=0}
,
\end{equation}
and
\begin{equation}
0 = 2 E r_p \Big( E\left(r_p^2 + a^2 \right)  - a L \Big) - K (r_p - m) 
\label{eq:R'+geod+func=0}
.
\end{equation}
From these one obtains
\begin{equation}\label{eq:division}
\frac{\Delta(r_p)}{(r_p - m)}
=
\frac{\Big( E\left(r_p^2 + a^2 \right)  - a L \Big)}{2 E r_p}
;
\end{equation}
where we have divided the above equations.
Note that $E$ is of order one and negative, while
$|aL| << |E r_p^2|$; implying that the $\Big( E\left(r_p^2 + a^2 \right)  - a L \Big)$
is negative, so that \eqref{eq:division} has the correct sign. 
Then, one has
\begin{equation}\label{eq:Lder_p}
L = 
\frac{1}{a}
\Big(
E\left(r_p^2 + a^2 \right)
-
\frac{2 E r_p\Delta(r_p)}{(r_p - m)}
\Big)
;
\end{equation}
and
\begin{equation}\label{eq:Kder_p}
K = \frac{(2 E r_p)^2 \Delta(r_p) }{(r_p - m)^2}
.
\end{equation}

One also must satisfy
\begin{equation}\label{eq:Titamayor0}
K - \left( \frac{L}{\sin(\theta_p)} - a E \sin(\theta_p) \right)^2 
\geqslant 0 .
\end{equation}

Then, \eqref{eq:Lder_p} gives $L(r_p)$, while \eqref{eq:Kder_p} gives $K(r_p)$.
To obtain the silhouette one must consider those geodesics that touch
the outer boundary of this regions, which is characterized by
the condition that $\dot{\theta_p}=0$ or equivalently
\begin{equation}\label{eq:Tita=0}
K(r_p) - \left( \frac{L(r_p)}{\sin(\theta_p)} - a E \sin(\theta_p) \right)^2 
= 0 .
\end{equation}
This allows to express $r_p(\theta_p)$, since one has
\begin{equation}\label{key}
\frac{(2 E r_p \sin(\theta_p) )^2 \Delta(r_p) }{(r_p - m)^2}
=
\left( L(r_p) - a E \sin(\theta_p)^2 \right)^2
,
\end{equation}
or
\begin{equation}\label{key}
\begin{split}
\Delta(r_p) &
\frac{(2 E r_p \sin(\theta_p) )^2  }{(r_p - m)^2}
= \\
&
\left(
\frac{1}{a}
\Big(
E\left(r_p^2 + a^2 \right)
-
\frac{2 E r_p\Delta(r_p)}{(r_p - m)}
\Big)
- a E \sin(\theta_p)^2 
\right)^2
,
\end{split}
\end{equation}
so that
\begin{equation}\label{key}
\begin{split}
\Delta(r_p) &
\frac{(2 a r_p \sin(\theta_p) )^2  }{(r_p - m)^2}
= \\
&
\left(
r_p^2 + a^2 
-
\frac{2  r_p\Delta(r_p)}{(r_p - m)}
- a^2  \sin(\theta_p)^2 
\right)^2 = \\
&
\left(
r_p^2 + a^2 \cos(\theta_p)^2
-
\frac{2  r_p\Delta(r_p)}{(r_p - m)} 
\right)^2 
.
\end{split}
\end{equation}
Taking the square root one would have a $\pm$ relation, but we can build
this in the definition of $\vartheta_p$ with a range of $[-\pi,\pi]$;
so that we can write
\begin{equation}\label{key}
\begin{split}
\sqrt{\Delta(r_p)}
&
2 a r_p \sin(\vartheta_p) 
= \\
&
(r_p - m)
\big(
r_p^2 + a^2 \cos(\vartheta_p)^2
\big)
- 2  r_p\Delta(r_p) 
;
\end{split}
\end{equation}
which provides for us $r_p(\vartheta_p)$.
Defining $f(r_p)$ as
\begin{equation}\label{eq:efe1}
\begin{split}
f(r_p) =&
(r_p - m)
\big(
r_p^2 + a^2 \cos(\vartheta_p)^2
\big)
- 2  r_p\Delta(r_p) \\
&-
\sqrt{\Delta(r_p)}
2 a r_p \sin(\vartheta_p) 
;
\end{split}
\end{equation}
one could solve for $r_p$ using the standard techniques of root finding.

Then we obtain $L(\vartheta_p)$ and $K(\vartheta_p)$;
from which we draw the silhouette using celestial coordinates.

Note that when $a=0$ one has 
\begin{equation}\label{eq:efe0}
\begin{split}
f(r_p) =&
(r_p - m)
r_p^2
- 2  r_p\Delta(r_p) \\
=&
r_p^3 - m r_p^2
- 2  r_p(r_p^2 - 2 m r_p) \\
=&
- r_p^3 + 3 m r_p^2
;
\end{split}
\end{equation}
so that $r_p = 3 m$ in this case.

It should probably be emphasized that
the silhouette shown in Fig.\ref{fig:image-with-silhouette}
is completely consistent with our calculations,
since all the photon trajectories are outside
of the silhouette, and the brightness
inside is just an effect that comes from the
fact that we are blurring the images with a
circular Gaussian filter, that allows to compare
our images with those from EHT;
as explained in detail in our previous
article\citep{Boero:2021}.

\section{Final comments}\label{sec:Final+commnets}

As commented previously, the EHT collaboration
has used as guiding idea, the structure of rings
for the construction of their images.
The question of whether Sgr~A${}^*$
is appropriately described by a ring, is so
central to the EHT 
work\citep{EventHorizonTelescope:2022wok}
that they dedicate a subsection to this point.
They affirm there that:
``there are a small number of
nonring images that ﬁt the data well 
and cannot easily be
excluded through additional tests.''

In our work we have applied as guiding idea the
structure of a disk, which was very successful
in the construction of images for the M87 system.
Since they also consider the width of the rings,
one may wonder whether there could be some
kind of intersection between the two main ideas
of the models. We can point out, that since
the main starting physical model are rather
different, also the language is different;
but it should be remarked that they
concentrate on rings that are mostly face-on,
in contrast to our assumption of inclined 
thin accretion disks,
that are mostly edge-on.

The main EHT image shown in Fig. \ref{fig:eht-image}
shows a basic structure of three local intensity
maxima located approximately at position angles
of: (A: 70$^\circ$),
(B: 220$^\circ$) and (C: 330$^\circ$);
on a disk like shape.
Note that east is to the left, and position angles 
are defined east of north.
Interpreting this configuration as a possible
disk, which was inclined to the observer,
and assuming that structure C indicated the
opposite side of the disk, we first carried out
a series of images, for the iota angle
in the range $\iota \in [-10^\circ, -80^\circ$],
and with the angular parameter $a$
with values $a=[0, 0.25, 0.50, 0.75, 0.98]$.
We conclude that the images so constructed
do not represent consistently the EHT image
of Sgr~A${}^*$.
For this reason we next performed calculation
of images that assume the more natural configuration
of a disk, with small angle variations with
respect of the plane of the galaxy.
In all cases we have assumed the disk is located
in the equatorial plane of the black hole,
and therefore, its angular momentum
is perpendicular to the disk.
With this second set of images, we find that
one can explain the structure B of the EHT
image of Fig. \ref{fig:eht-image}.
The model we have used does not account for
the other structures A and C of the EHT image.

As it has been mentioned in the EHT articles,
the acquisition of the data and its processing is
a very difficult task that forces the team to
use models, with different weights, in the
reconstruction of images.
In particular they mention in
reference \citep{EventHorizonTelescope:2022wok}
the use of eight synthetic models for
their image reconstruction;
from which six of them show
a face-on orientation,
three ring variants, two disk
variants and a GRMHD model.

It should be remembered that the final EHT
image is the average of four different images
that they show in their Figure 13 graphs.
However it is noticed that only structure B
is persistent in the four basic images;
probably indicating that it is a stable
physical structure observed by the EHT team.

\begin{figure*}
\centering
\includegraphics[clip,width=0.7\textwidth]{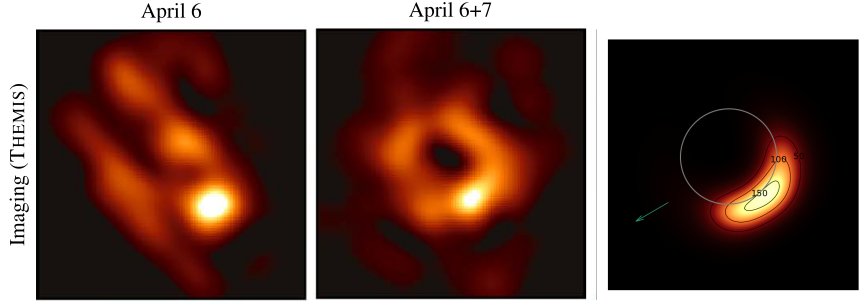}
\caption{Reproduction of two graphs first row of Figure 27 of
paper IV of the EHT work on Sgr~A${}^*$; they correspond to
April 6 and April 6+7 of the {\sc Themis} pipeline.
The third graph on the right is the copy of our image
at the scale of the other EHT images.
}
\label{fig:de-fig-27-paper-iv}
\end{figure*}

This fact is reinforced if we take a look
at the first row in Figure 27
of the EHT Collaboration article
\cite{EventHorizonTelescope:2022exc}
which we reproduce here in
the left two graphs of 
Fig.
\ref{fig:de-fig-27-paper-iv}.
The relevant graphs, that we reproduce
in Fig.
\ref{fig:de-fig-27-paper-iv}
are the first and the fourth of their Figure 27;
where they show the images that
are obtained from the 
{\sc Themis} image algorithm.
In their paper III, in which they
present the preferred image of
Sgr~A${}^*$, they have chosen
to use the data of April 7.
However, it is noticeable that
the {\sc Themis} image from
April 6, shows a prominent
local intensity maximum more or 
less at the position of the
local intensity maximum of our preferred
image shown in Fig. 
\ref{fig:image-with-silhouette}.
Then, although they do not trust the April 6
{\sc Themis} image, we instead think that
this gives support to our conjecture
that structure B corresponds
to an intrinsic persistent structure 
of Sgr~A${}^*$.

To help the comparison of our work with the EHT work,
we show in Fig. \ref{fig:de-fig-27-paper-iv}
the two EHT images from the {\sc Themis} pipeline
for April 6 and April 6+7, and our best choice image,
which  contains the silhouette and the direction
of the angular momentum, using the same
angular scale.
We conclude that they compare successfully, since 
we reproduce very well the position of the brighter
part of the images.

We hope that our work can contribute to the
adjustments of pipelines that are used in those
reconstructions processes; since different
priors produce various outcomes.

\subsection*{Data Availability}
No new data were generated or analysed in support of this research.
The numerical calculation is completely described in the article.

\subsection*{Acknowledgments}

We are grateful to the EHT Collaboration
for publishing their images
under the terms of the
Creative Commons Attribution 4.0 licence.

We acknowledge support from CONICET, SeCyT-UNC and Foncyt.


\bibliographystyle{mnras}

\bsp	
\label{lastpage}
\end{document}